# Detection and Prevention of Smishing Attacks

## A DISSERTATION

*submitted in partial fulfilment of the
requirements for the award of the degree
of*

**Master of Technology**

*in*

COMPUTER ENGINEERING (CYBER SECURITY)

**by**

DIKSHA GOEL
Roll No.: 31603217

*Under the supervision of*
**MR. ANKIT KUMAR JAIN**

Assistant Professor

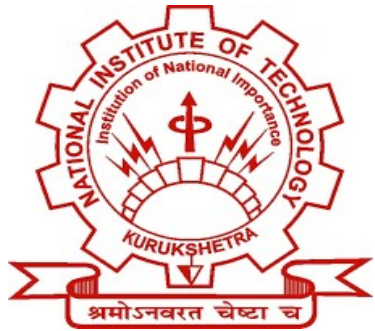

**DEPARTMENT OF COMPUTER ENGINEERING
NATIONAL INSTITUTE OF TECHNOLOGY
KURUKSHETRA-136119, HARYANA (INDIA)**

**JUNE, 2018**

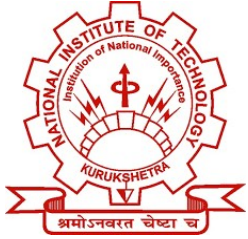

# CERTIFICATE

I hereby certify that the work which is being presented in this M. Tech. Dissertation entitled **"Detection and Prevention of Smishing Attacks",** in partial fulfilment of the requirements for the award of the **Master of Technology in Computer Engineering (Cyber Security)** is an authentic record of my own work carried out during a period from June 2017 to June 2018 under the supervision of **Mr. Ankit Kumar Jain, Assistant Professor,** Computer Engineering Department.

The matter presented in this thesis has not been submitted for the award of any other degree elsewhere.

*Signature of Candidate*
**DIKSHA GOEL**
**Roll No. 31603217**

This is to certify that the above statement made by the candidate is correct to the best of my knowledge.

*Signature of Supervisor*

**Date:**
**Place:** Kurukshetra

**Mr. Ankit Kumar Jain**
**Assistant Professor**
**Department of Computer Engineering**
**National Institute of Technology**
**Kurukshetra-136119 (Haryana)**

# ACKNOWLEDGEMENT

I came across a number of people whose contribution in different ways helped me during my research work and while bringing out this dissertation to its final form. It is a pleasure for me to express them a sincere appreciation and a special thanks.

I would like to express my deepest gratitude to my dissertation supervisor, **Mr. Ankit Kumar Jain**, for his excellent guidance, caring, patience, and providing me with an excellent research atmosphere. I thank him for all the time he spent with me discussing everything from my research work to career options, his supervision has helped me during setbacks and obstacles and inspired me to give my best. I would like to thank **Dr. B. B. Gupta,** Assistant Professor, Department of Computer Engineering for his valuable suggestions in the field of information and cyber security, and practical issues beyond the textbooks.

I am highly grateful to **Dr. R. K. Aggarwal,** Head of Department of Computer Engineering and **Dr. A. K. Singh,** Former Head of Department of Computer Engineering for their assistance during the entire course. I would like to thank other faculty members and non-teaching staff of the department for their valuable help and feedback.

I am grateful to my parents and younger brother for being supportive and encouraging me with their love and blessings. My research would not have been possible without their help. I am thankful to all my friends for supporting me and motivating me by providing their valuable recommendations. Finally, I would like to thank all the dignitaries who have been involved directly or indirectly with the completion of my work.

<div style="text-align: right;">**Diksha Goel**</div>

# ABSTRACT

Phishing is an online identity theft in which an attacker tries to steal user's personal information, resulting in financial loss of individuals as well as organizations. Now-a-days, mobile devices especially smartphones are increasingly being used by the users due to a wide range of functionalities they provide. These devices are very compact and provide functionalities similar to those of desktop computers due to which attackers are now targeting the mobile device users. However, detection of mobile phishing attack is a different problem from desktop phishing due to the dissimilar architectures of both.

Smishing attack refers to performing phishing attack using Short Messaging Service (SMS). It is SMS based identity theft in which an attacker sends an SMS aiming to steal personal information of the mobile device user. With the increase in usage of SMS-based services, smishing attacks have also increased. Due to an exponential increase in the smishing attacks from the past few years, it is very important to address these attacks. Although various techniques are available for detecting spam messages, but very less amount of work has been done in the field of detection of smishing messages.

In this dissertation work, we propose a smishing security model based on Content-based analysis approach. Now-a-days slang words, abbreviations and short forms are commonly used by the users in text-based communication that somehow reduce the efficiency of the classifiers. To address this limitation, we normalize the short forms into their standard form. Machine learning classifier is used to classify the message as smishing message or ham message. We evaluated our approach using dataset and the results from the experimental analysis show that our model gives 97.14% classification accuracy in smishing messages and 96.12% classification accuracy in ham messages, achieving an overall classification accuracy of 96.20%.

# Table of Contents

# List of Figures

# List of Tables

# List of Abbreviations

| Abbreviation | Description |
| --- | --- |
| SMS | Short Messaging Service |
| URL | Uniform Resource Locator |
| DDoS | Distributed Denial of Service |
| APWG | Anti-Phishing Working Group |
| OS | Operating System |
| AOL | America Online |
| VoIP | Voice over Internet protocol |
| UI | User Interface |
| LUI | Login User Interface |
| XSS | Cross Site Scripting |
| SSID | Service Set Identifier |
| MITM | Man in the middle |
| SSL | Secure Socket Layer |
| SID | Session Id |
| DNS | Domain Name Server |
| SVM | Support Vector Machine |
| DCA | Dendritic Cell Algorithm |
| LIWC | Linguistic Inquiry and Word Count |
| SMSS | SMS Specific |
| IG | Information Gain |
| OCR | Optical character recognition |
| SLD | Second Level Domain |
| TCP/IP | Transmission Control Protocol/ Internet Protocol |
| VPN | Virtual Private Network |
| HTTPS | Hyper Text Transfer Protocol Secure |
| QR | Quick Response |
| TP | True Positive |
| FP | False Positive |
| TN | True Negative |
| FN | False Negative |
| TPR | True Positive Rate |
| FPR | False Positive Rate |
| TNR | True Negative Rate |
| FNR | False Negative Rate |
| NSC | NUS SMS Corpus |
| NLP | Natural Language Processing |
| CSV | Comma Separated Values |
| NLTK | Natural Language Toolkit |

# CHAPTER 1

# INTRODUCTION

With the advancements in technology, various smart devices such as desktop, smartphones and tablets are being developed. Smartphones have become an integral part of our everyday life. These are not only small in size and easy to carry, but also provide functionalities similar to those of desktop computers. With the increase in usage of smartphones, cyber threats related to smartphones have also increased. Phishing is one such cyber threat. Phishing attack refers to an act of stealing sensitive credentials of the users. It is also used for installing malicious software on victim's device. Phishing attacks help attackers in performing various illegal activities like fraud and identity theft.

Recent market trends have made attackers to shift their focus from desktop users to mobile device users. SMS is one of the most trustworthy text based communication channel used by users and is widely used by attackers for carrying out phishing attack. Smishing is SMS based Phishing, social engineering attack where assailant sends a phishing SMS to the users with the intention to steal their sensitive information. A lot of solutions have been proposed for the detection and prevention of phishing attacks, still the threat is not alleviated. Blacklisting, Uniform Resource Locator (URL) based detection, static detection, and heuristics methods are various techniques used for detecting phishing attacks. However, whenever researchers come with some ideas to control phishing attack, attackers change their attack strategy and exploit vulnerabilities of the current solutions. In this chapter, we provide an overview of phishing attack and smishing attack, and the motivation to work on the same. Further, we discuss the objectives of the dissertation and lastly, organization of dissertation is presented.

## 1.1 Overview

Cyber threats such as unsolicited emails, malicious software, viruses, spyware, Distributed Denial of Service (DDoS) attacks, and social engineering attacks exploit the security of smartphone devices. One such cyber threat is "Phishing Attack". Phishing attack is an online spoofing mechanism in which social engineering messages are communicated via electronic communication channels to prompt users to perform certain actions for the benefit of attacker [1]. Attackers launch phishing attack for social or financial gains. A recent report by Anti-Phishing Working Group (APWG) shows that in 2016, the total number of unique phishing attacks detected were 1,220,523, which is an increase of 65% over 2015 [2-5]. The real target of phishing attack is not the infrastructure, but the users. Phishing is a complex issue and is continuously changing its ways towards the target victims. Now-a-days, attackers are using Trojan, virus and ransomware along with phishing attacks in order to exploit the vulnerabilities of the smartphone devices.

Use of smartphones have become prevalent due to their small size, long battery life, and portability [6]. Easy availability of smartphones and low-cost data plans have led to an exponential increase in the usage of smartphones. Not only young or techno-freak people are using smartphones, but these are used by people of all age groups. With an increase in usage of smartphones, security threats related to these devices have also increased. Smartphone devices have become an attractive target for the attackers. Attackers may launch phishing attack over mobile phone through various mediums, such as SMSes (smishing) [7], VoIP (vishing) [8], emails (spear phishing, whaling) [9], web browsers [10], mobile applications [11] and social networking websites [12]. Out of these services, SMS is widely used to launch phishing attacks. With more than 2 billion smartphone users across the world, 20 billion text messages are sent everyday with an average opening time of less than 3 seconds [13]. These huge figures have attracted attackers to launch phishing attacks using SMS, also known as smishing attack.

The term smishing is composed of two words "SMS" and "Phishing". Smishing messages are subset of spam messages. Smishing attack is highly targeted phishing attack. Instead of sending phishing SMS to any random user, an attacker sends phishing SMS to those who appear to be an attractive target to them [14]. From the past few years, these attacks have affected the security and privacy of the mobile device users. Sometimes attacker sends a link to malicious application or a phishing webpage to the users via SMS. Various malware may enter the device through smishing attacks. Attackers are taking advantage of trust and dependency of users on their smartphones. Due to an exponential increase in the smishing attacks from the past few years, it has become a major concern to address these attacks. Detection of smishing attack is a big challenge faced by the researchers and it requires solution that can identify smishing messages with higher accuracy.

Large number of phishing websites revive and expire every day. According to APWG report, on an average, a phishing site stays on the web for 4.5 days and sometimes for just few hours [15]. Security of mobile devices is influenced by many factors, such as security threats and security requirements. We do not have any method to check if the credentials are sent to a legitimate server or any other rogue server. By exploiting the hardware limitations of mobile phone devices and careless behavior of the users, an attacker can easily carry out phishing attack on mobile phones. If Operating System (OS) of mobile devices is compromised, malicious applications can access device's camera, SMSes, contacts, and can also gain location information which in turn compromise the privacy of the users [16]. There is a lack of knowledge among the users about phishing attacks and how they can be avoided [17]. According to a study, 44% of the users are not aware of the security solutions available for mobile devices [18]. Economy is negatively affected by the phishing attacks due to financial losses that are faced by the businesses as well as the individuals [19].

## 1.2 Motivation

Smartphones have attracted a large number of users due to wide range of functionalities they provide. Smartphones are not only used for messaging, making phone calls, gaming, but also for performing other tasks such as financial transactions, business inquiries, subscriptions, browsing and online shopping. Smartphone users are increasing at an exponential rate and are expected to be 2.87 billion by the end of year 2020 [20]. Smartphone users are three times more vulnerable to phishing attacks as compared to desktop users [21] and the reason for these vulnerabilities are - small screen size, inconvenience for the users to input, switching between applications, lack of security indicators, and habits and preferences of mobile device users.

With the increase in usage of smartphone, smishing attacks are also increasing. One out of every three mobile phone user has received smishing messages [22]. Now-a-days, it has become very easy to carry out phishing attack using SMS as low cost SMS plan enable attackers to send phishing messages to large number of users. A report by Wombat states that in 2017, 45% of the users have received smishing messages which is an increase of 2% from 2016 [23]. Phishing SMS generally contains a message along with a malicious link. Studies have shown that 42% of the mobile device users click on the malicious link [24]. In 2016, Edward Smith, a client of UK bank Santander lost £22,700 in a Smishing scam [25]. Smishing attack targets the sensitive information of the user.

Hence, it is very important to detect and prevent these attacks. There are many solutions available for detecting spam messages, but not much amount of work has been done to detect smishing messages. Therefore, it is important to come up with a solution that can efficiently detect smishing attack with high accuracy.

## 1.3 Dissertation Objective

Smishing messages are the subset of Spam messages. Spam messages are unsolicited messages such as subscription messages, offers, sales and advertisements, while Smishing messages are the spam messages indented to steal sensitive information of the users. The main objective of our dissertation is to design a smishing security model that can detect smishing messages with higher accuracy. However, the task is challenging as slang words, abbreviations and short forms are commonly used by the users in text-based communication that somehow reduce the efficiency of the classifiers. The key factor to be considered while dealing with SMSes is to maintain the privacy of the users. Thus, in this dissertation work, our objective is to design a solution that effectively detects and block smishing messages, while maintaining the privacy of the users.

## 1.4 Dissertation Organization

The organization of the remaining part of the dissertation is as follows –

In **Chapter 2**, we provide comprehensive details about phishing attack and smishing attack including its historical background and some statistics. We also present taxonomy of phishing attacks in mobile computing environment and some existing defence mechanism.

In **Chapter 3**, we describe our proposed smishing security model in detail which includes working of different phases.

In **Chapter 4**, we discuss the performance metrics and experimental setup i.e. tools and dataset used, and the results of the proposed solution. We also present a comparative analysis of the proposed approach with the other existing approaches.

In **Chapter 5**, we conclude the dissertation work and provide insights about the future research that can be done in this area.

# CHAPTER 2

# LITERATURE REVIEW

Phishing attack has become one of the most serious cyber threat. Attacker targets the vulnerabilities found in users rather than that of system via phishing attack as users can technically secure their systems by using various logins, antivirus and plug-ins but what if users themselves reveal their passwords. Various attacking mechanisms are used by the attackers to launch phishing attack such as social engineering, mobile applications, web browsers and social networking sites etc. In this chapter, we provide comprehensive details about Mobile phishing attacks including its historical background and some recent statistics. We also discuss life cycle of mobile phishing attack. We analyze various mobile phishing attacks and propose a taxonomy of mobile phishing attacks. We also present a taxonomy of numerous recently proposed solutions that detect and defend users from mobile phishing attacks. In addition, we outline various research issues and challenges associated with mobile phishing attack that needs to be addressed.

## 2.1 History and Background

The word phishing comes from the word 'fishing'. As a fisherman uses his bait to catch fishes, similarly the attackers use social engineering messages to get sensitive information from the users. The reason behind using "ph" in place of "f" in the term phishing is that earlier hacker used "ph" to mean "phone phreaking" which was one of the earliest forms of hacking. The term "Phishing" was coined in the year 1996, when a large number of fraudulent users with fake credit card details registered on America Online (AOL) website. AOL approved these accounts without verification and attackers started using AOL system's resources. At the time of payment for the services, AOL found that most of these credit cards were invalid and accounts were fake. As a consequence, the accounts were ceased. After this, AOL started properly authenticating the credit cards. This made attackers to find other ways for obtaining AOL accounts. After that, instead of using fake accounts, attackers started stealing passwords of registered AOL users by contacting them through emails or messages that appeared to be from AOL employees and used various services on behalf of the legitimate users by using their credentials [26]. First incident of smishing attack was tracked in 2004 and it increased significantly in 2006. Table 2.1 outlines some of the major events associated with the evolution of phishing attack.

Table 2.1 Evolution of Phishing Attack

| Year | Events |
| --- | --- |
| 1996 | "Phishing" term was used for the first time |
| 1997 | Media warned users about Phishing attacks |
| 2000 | Use of key loggers for obtaining credentials |
| 2001 | Use of spam messages for phishing attacks |
| 2003 | Attackers started registering for domains that mimic like legitimate sites |
| 2004 | "Cabir", the first mobile malware was released |
|  | Use of SMS to conduct phishing attacks |
| 2005 | Use of Spear Phishing attacks |
| 2006 | Use of "Man in the Middle" attack for Phishing attack; |
|  | Use of VoIP to conduct phishing attacks, also known as Vishing |
| 2008 | Mobile Application Store was launched |
| 2009 | "Chat in the Middle" Phishing attack was discovered |
| 2010 | Term "Tabnabbing" was coined |
| 2011 | Gaming users hit by "Xbox Live" phishing attack |
| 2014 | 110 million credit cards were compromised due to phishing scams |
| 2015 | 100,000 people received phishing emails in UK |
| 2016 | 500% increase in social media phishing attacks |
| 2017 | W-2 Phishing attacks affected more than 120,000 people |

## 2.2 Statistics of Mobile Phishing Attacks

Capabilities of personal computers are combined with the pocket sized mobile phones resulting into a device called Smartphone. Smartphones give users a wide range of functionalities, such as calling, SMS, emails, downloading, gaming, audio and video playback. These rich functionalities have attracted a large number of users [27]. Smartphone users are increasing every year. Currently there are 2.32 billion smartphone users worldwide which is expected to be 2.87 billion in 2020 as shown in Figure 2.1 [28]. Mobile traffic exceeded desktop traffic for the first time in November 2016, showing a market swing about how people are currently accessing the web. Attackers are very much aware about this shift and hence targeting the mobile device users instead of desktop users.

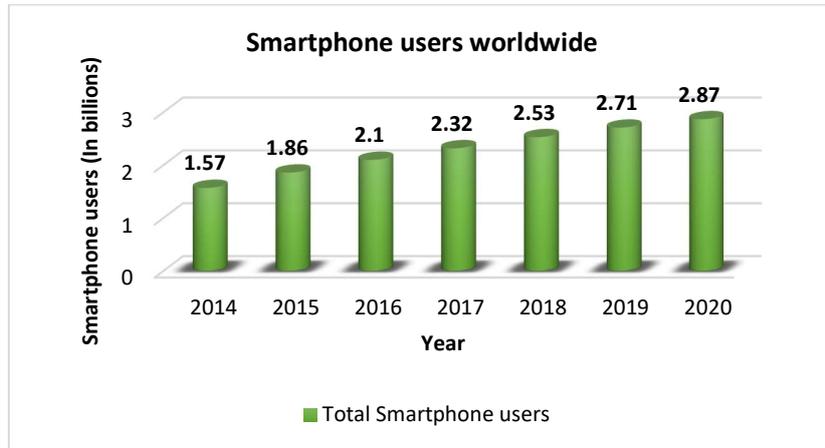

Figure 2.1: Number of smartphone users around the world

Attackers employ various techniques such as creating phishing websites, sending phishing emails, and SMSes, and trick users into revealing their personal information. Now-a-days, attackers are using SMSes to perform phishing attacks over mobile phones. SMS is one of the most widely used and trustworthy text based communication channel used by the users. According to Portio Research report [24], SMS traffic increased to 100 billion in just 3 years from 1996 to 1999. By 2003, SMS traffic reached to 450 billion messages. In 2009, this traffic crossed 5 trillion and in 2015 it reached to 8.3 trillion as shown in Figure 2.2.

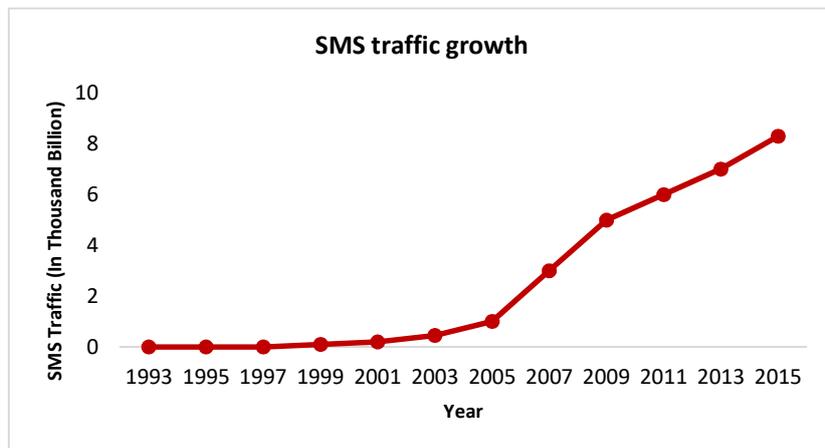

Figure 2.2: SMS traffic growth

Figure 2.3 illustrates trust level of users on different messaging platforms. 35% of the users consider SMS as one of the most trustworthy channel, 28% of the users trust various messaging applications like WhatsApp, Instagram, while 18% of the users trust messenger, Facebook, Yahoo and Skype [29].

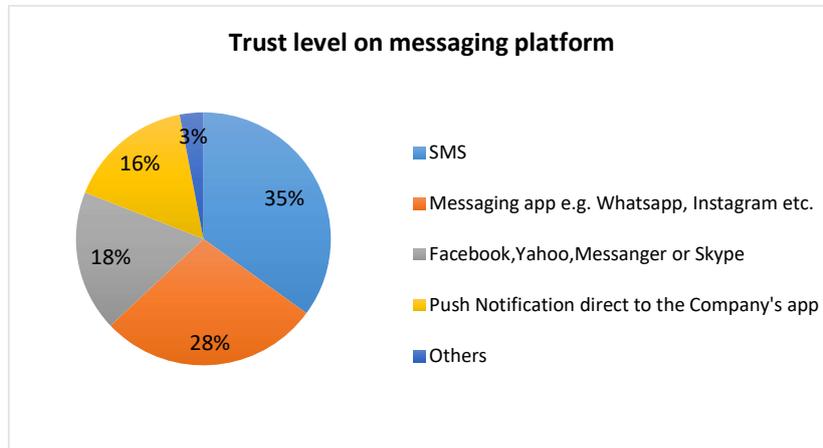

Figure 2.3: Trust level of user on various messaging platforms

Smishing attack is an SMS based phishing attack where user is tricked into downloading viruses, worms, or other malware. Smishing attack is a variant of email phishing and generally considered as a regional consumer based threat rather than global cyber security concern [23]. Attackers prefer SMSes over other communication mediums such as phone calls, Facebook, or emails as the response rate of SMS is 209% higher than that of these mediums [30]. In addition, only 17% of emails are opened whereas 99% of SMSes are opened within few minutes after being received by the user [31]. A report reveals that 33% of all mobile phone users have received a smishing message offering various deals and discounts [32]. A survey conducted by Wombat [23] shows that most of the users do not know about smishing attacks. Only 16% of the users gave right answer when asked "what is smishing", 17% users gave wrong answer, whereas majority of the users i.e. 67% users said that they do not have any idea about smishing attacks [23].

According to Cloudmark threat report, 25% of spam messages are smishing messages [33]. Due to lack of awareness among users, most of the users on receiving these unwanted messages either ignore them or do not take any actions as shown in Figure 2.4. Studies have shown that 54% of the users delete these messages, 17% of the users ignore them, 13% of the users answers with stop command and only 16% of the users report them [29]. Most of the users are not aware that phishing attack can be performed via text message which in turn increases the chances of users falling victim to smishing attacks. According to PhishingPro report 2016 [34], earlier 90% of successful phishing attacks begin with a phishing email. Attackers used spear phishing to target the users. But now attackers are using mobile applications instead of emails to gain credentials. In mobile phones, 81% of the phishing attacks are carried out using mobile applications, SMS, or websites while only 19% of the phishing attacks are carried out using mobile emails [35].

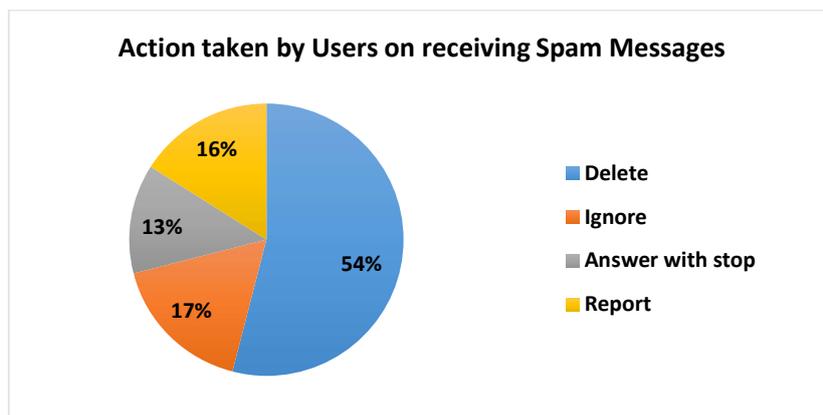

Figure 2.4: Action taken by user on receiving spam messages

## 2.3 Mobile Phishing Lifecycle

In phishing attack, the attacker tricks the user into revealing their sensitive information by pretending to be a trustworthy entity such as person, firm or any government organization. Figure 2.5 shows the lifecycle of phishing attack. The phishing attack can be performed as follows:

- *Planning phase* – In the first phase, the attacker selects a feasible communication media for initiating the phishing attack. This media can be a phishing webpage, a phishing application, an email, or an SMS containing a malicious link. The attacker sets the target that can be an individual, organisation, or a nation and collect details about them by physically visiting them or monitoring them. After that attacking technique is selected that can be malicious application, or SMS.
- *Phishing phase* – In this phase, material is propagated to victim. The attacker sends phishing material to mobile device users using spoofed SMSes or emails pretending to be a legitimate source.
- *Penetration phase* – Once the user opens the propagated material, either a login page appears which redirects user to a phishing webpage and asks for personal information, or a malicious application is downloaded, allowing the attacker to penetrate in the device.

- ***Data gathering phase –*** As soon as the attacker gets access to the device, user's information is extracted either through the malicious application or fake login page. If a malware is installed in the device then the attacker can get remote access to the device and can get whatever information he wants from the device. Attackers can use gained information for financial benefits or other purposes.

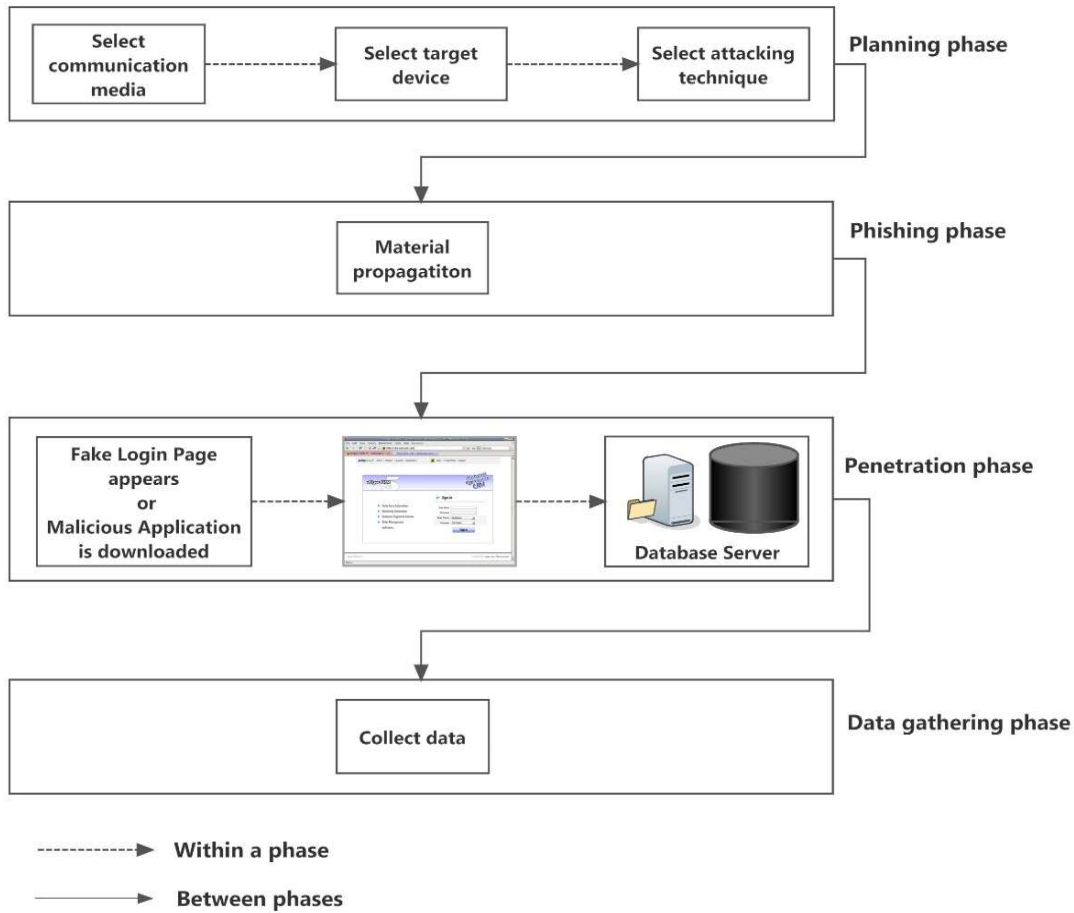

Figure 2.5: Mobile phishing life cycle

## 2.4 Taxonomy of Mobile Phishing Attacks

Mobile phishing attacks can be classified into various categories based on social engineering, mobile applications, malware, social networking sites, content injection techniques, or wireless mediums. The taxonomy of mobile phishing attacks is shown in Figure 2.6. Some of the mechanisms for carrying out phishing attacks on mobile phones are discussed below.

## 2.4.1 Phishing through Social Engineering

Social engineering is an art of deceiving users into disclosing their sensitive information. Instead of attacking the systems, social engineering attacks target the humans who have access to information and manipulate them into revealing their confidential information. In this attack, technical protection is not much effective as it targets the users and not their devices. In addition, people believe that they are smart enough and would not fall for such attacks [36]. Social engineering attacks take advantage of ignorance and careless behaviour of users due to which sensitive information may be revealed. There are various social engineering approaches such as physical approach, social approach, reverse social engineering, technical approach, and socio-technical approach [37]. Some of the social engineering methods for carrying out phishing attacks on mobile phones are discussed below.

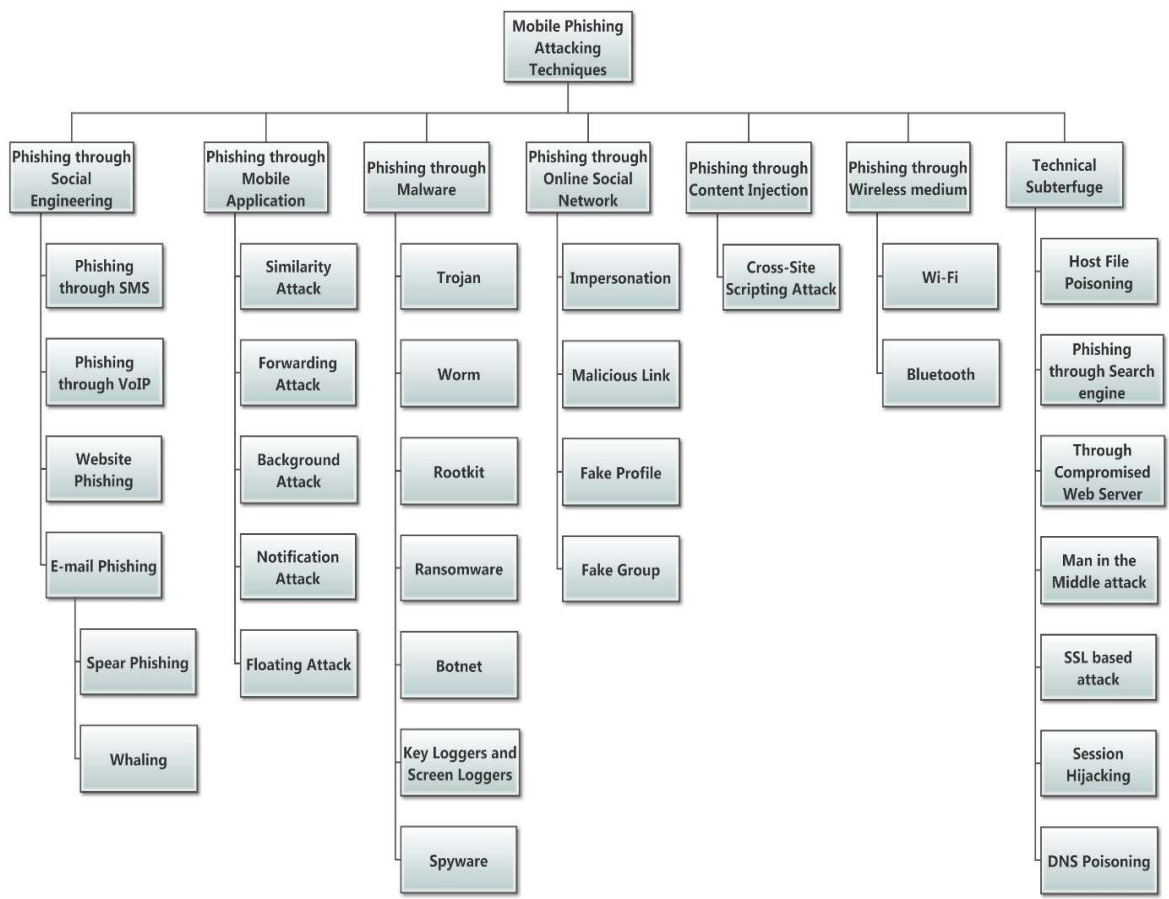

Figure 2.6: Taxonomy of mobile phishing attacks

## 2.4.1.1 SMS

One of the most popular methods to carry out phishing attacks on mobile phones is through SMS, and this method is called as Smishing. This attack is intended to steal personal and financial information over the mobile phones. Smishing messages usually contain text message along with a link which when opened either redirects user to a fake website, or some malicious program is installed [38]. Through smishing attack malware can enter the device. This attack is based on social engineering and users are easily targeted by it. Various methods have been proposed to detect malicious links included in the SMS but these links are changed frequently. Also URLs are shortened which makes it even more difficult to detect malicious URLs [7]. Various reports clearly state that smishing attacks have exponentially increased over the past few years. Smishing attacks can be carried out using a number of techniques. Figure 2.7 shows the example of the smishing messages and some of the techniques used by the attackers are discussed below [25].

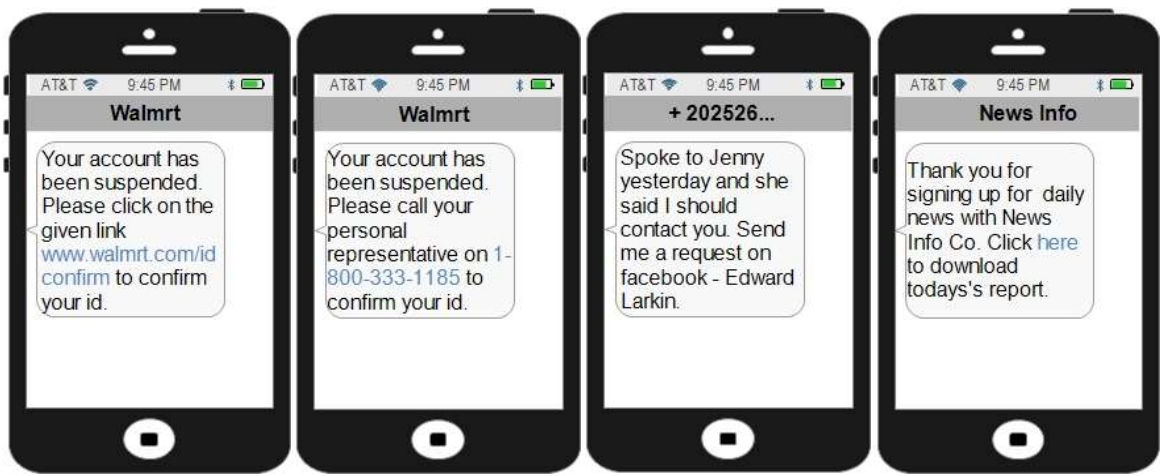

Figure 2.7: Example of Smishing Messages

- *Use of bogus links in the SMS* – In this type of smishing technique, attackers send SMS to the users pretending to be from a well know organization such as bank, asking the users to confirm their identity in the provided link. When the users click on the link, it redirects them to a phishing webpage and asks for their personal information.
- *Use of fake phone number in the SMS* – Attacker may send an SMS to the user asking them to confirm their identity on the given phone number. When the user makes a call to that phone number, it connects user with eloquent phisher who tries to extract sensitive information from the user. Since these days, call centres are the support hubs for most of the companies, so users do not doubt call centres consultants asking personal questions.

- ***By pretending to be known entity*** – An attacker may pretend to be a known person to the users and requests them to connect on social networking sites. Once connected, attacker tries to gather information about the users from the contacts and other sources available on the social networking websites.
- ***By making users download a malicious application*** – In this type of smishing attack technique, attacker sends a message to the users tricking them into clicking the malicious link, and when the users click on that link, a malicious application is downloaded in the device. This application sends user's personal information stored on the mobile phone to the attacker.

### 2.4.1.2 Voice over Internet Protocol (VoIP)

Phishing over VoIP [8] has become very common and is also known by the name Vishing. Vishing is same as that of phishing and is carried out over phones using voice technology. Attackers gather information about the users, such as name, phone number, address, bank details and this is the information a genuine caller is expected to have. Moreover, users act without thinking during the phone calls [39]. Attacker sets the target user as victim and make him reveal his personal details. Vishing has a comparatively higher success rate than other mobile phishing methodologies because trust level that is built over the phone during the call is greater than other methodologies relying on the Internet [40].

### 2.4.1.3 Website

In website phishing attack, attacker targets individuals instead of a system. It is very easy for an attacker to create an exact replica of a legitimate website. Attacker tries to trick users by creating a phishing website of some famous websites, such as eBay, PayPal to obtain user's financial and personal details. A phishing website can be a legitimate website with phishing content inserted into it, or it can be a website owned by the phisher [1]. Blacklisting and heuristics based detection methods are used to detect phishing webpages. Blacklist contains suspicious IP addresses and URLs. In this approach, we search for the suspicious website in the list. Although it provides low false positive rate but does not protect from the Zero-Day phishing attacks. Heuristics methods are based on features which are present in most of the phishing webpages but not in all webpages. To bypass this, the attacker may design a website which may not have these features at all.

**2.4.1.4 Email**

Phishing email is a kind of spam message. These emails are illegal and rely on fake claims to be originating from legitimate company. The phisher sends a large number of fake emails along with a malicious link to users, requesting them to update their information in the provided link which when clicked either redirects user to a phishing website instead of legitimate one or ends up downloading a virus. Most of the approaches used to identify phishing emails are based on supervised, unsupervised and hybrid learning [41]. Many techniques have been developed to detect phishing emails, but still there is a lack of complete solution. Sequence diagram of mobile phishing attack through email is shown in Figure 2.8.

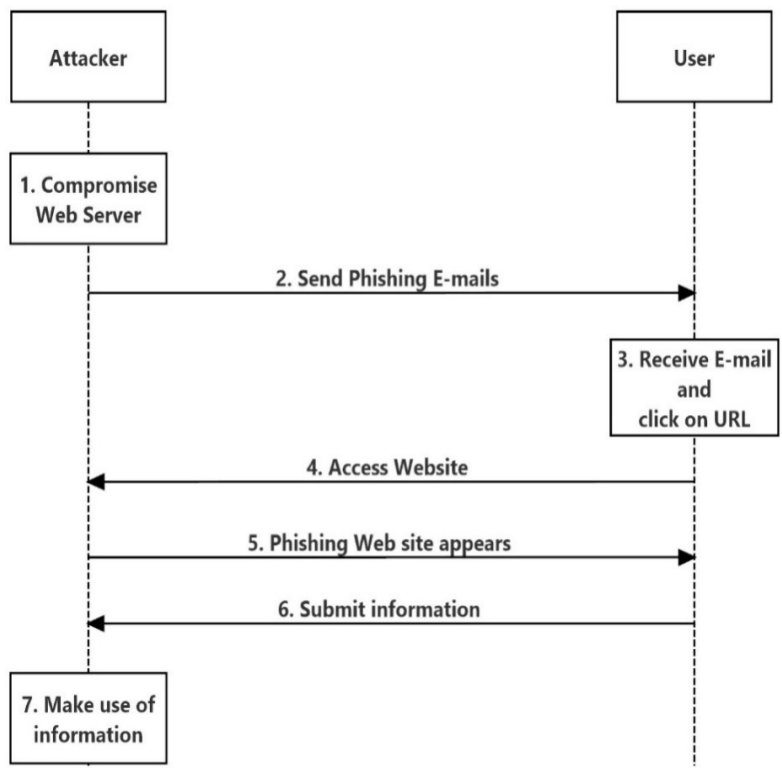

Figure 2.8: Sequence diagram of E-mail phishing

- ***Spear Phishing*** – A new type of email phishing is spear phishing in which an attacker targets an individual, organisation or a business group [42]. Spear phishing is effective as 70% of the users open these emails, and within an hour of receipt, 50% of them click on the link contained in the email [38]. In this attack, message appears to come from well-known organisations or companies. Spear phishing attack is successful if the source of the message appears to be trustworthy, the information contained in the message support its validity and requests that the message contains seems to be logical [43].

- *Whaling* – It is a kind of phishing scam that targets high profile companies or executives having valuable information. The target is chosen carefully according to the access and authority the person holds in the company. Attackers use social engineering to deceive users into disclosing their banking or personal details. Since these attack do not use malicious URLs or attachments, these attacks are difficult to detect [44]. Since the profit amount that the attackers obtain from this attack is large, they spend comparatively more time and effort to make this attack successful [45].

**2.4.2 Phishing through Mobile Applications**

Application based phishing attacks are a major problem on the mobile devices. During browsing or downloading an application, a user may fall victim of phishing attacks. Once malicious applications enter the device, they collect personal information of user like login ids, passwords and transmit the same to the attacker. Attacker may install a backdoor and other application which can breach privacy of the user [46]. Figure 2.9 shows the process of distribution of malicious applications in mobile phones. Small screen size and lack of security indicators makes it difficult to detect phishing attacks on mobile devices. Mobile application oriented phishing attack is classified into two categories. First, when phishing application tries to hijack existing legitimate application. Second, when phishing application directly pretends to be a legitimate application. This happens when user downloads a fake application from an unauthorized application market [47]. During installation, the application asks for various permissions that does not go with what the application is supposed to do [48]. Various phishing attacking techniques on mobile application are discussed below.

- *Similarity attacks* – In similarity attack [49], the phishing application, webpage or login interface have the same name, User interface (UI), and icon as that of the legitimate one. The attacker prompts the user to install phishing application and give login information in phishing Login User Interface (LUI) instead of legitimate one.

- *Forwarding attacks* – Forwarding attack is a type of phishing technique. In this attack, a phishing resource encourages users to share his activities like high score in a game on social networking sites and requests to launch the social networking application. The user when clicks the button to launch the social networking application then instead of launching the social networking application, a phishing login page is displayed. The phishing page asks for the login credentials to access the account. These types of forwarding attacks are difficult to detect.

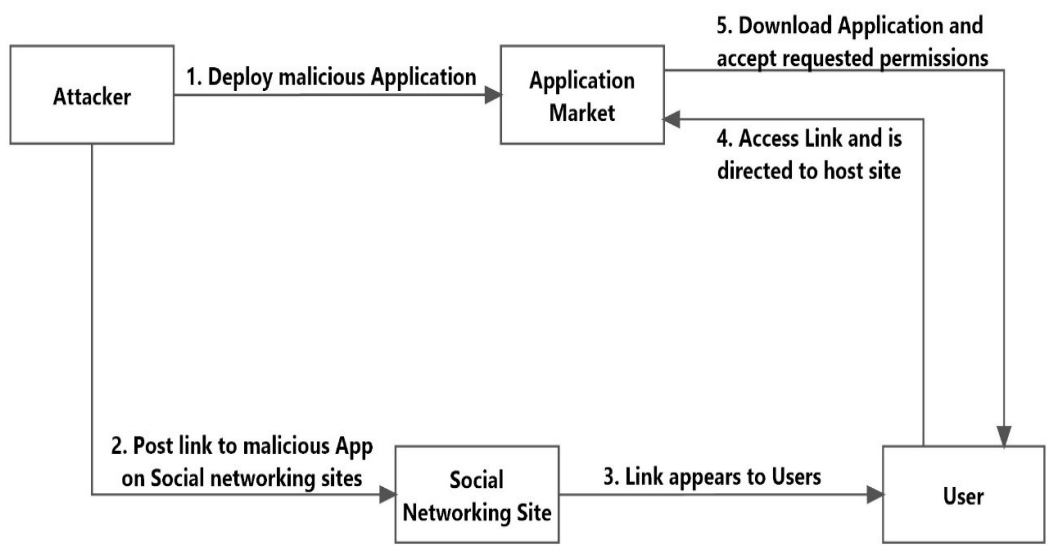

Figure 2.9: Distribution of Malicious Applications in Mobile phones

- *Background attacks* – Sometimes the malware or the phishing application runs and wait in the background and use ActivityManager of Android, to keep track of other applications running on the device. Whenever the user initiates a genuine target application, the phishing application turns-on itself in the foreground, and phishing screen is displayed.
- *Notification attacks* – The attacker might show a false notification asking the user for the personal details. The notification window can be modified by the attacker to look exactly like the genuine notification window.
- *Floating attacks* – The attacker uses the feature of Android device that permits an application to draw some action on top of that application in the foreground. Phishing application having the SYSTEM ALERT WINDOW permission may show a see-through input field over the login id and the password input field of the genuine application [49]. The LUI of the genuine application is seen by the user, but the coated input field is not visible to the user. Whenever the user enters the credentials in the input field, credentials are received by the phishing application.

### 2.4.3 Phishing through Mobile Malware

The term "Malware" was first used in 1990, by computer scientist and researcher Yisrael Radai [43]. Malware is a malicious software that gains access to device of the user without his consent. It is designed to steal data from the device, damage the device, or to annoy the user. Malware spreads via malicious attachments or malicious links. Attacker deceives user into installing malicious application and gain unauthorized access to the user's device. Malware secretly access personal data of the users and send it to its host or attacker. Malware takes advantage of loopholes present in the operating system or browsers, to deceive users and encourage them to execute the code [50]. In 2015, 430 million new unique malware were discovered by Symantec [18]. Table 2.2 shows various smartphone attacks, their effects and possible solutions.

Table 2.2 Smartphone attack, their effects and solutions

| Attack | Effect | Solution |
|--------|--------|----------|
| Trojan | Collect confidential information | Anti-virus and device specific intrusion detection system |
| Worm | Create backdoor for the attacker | Antivirus |
| Rootkit | Hide worms, bots and malware | Update device patches |
| Spyware | Collect confidential information | Firewall, Anti-spyware |
| Virus | Unusual behaviour of device | Antivirus |
| Backdoor | Effect security of device | Anti-virus and update device |
| Malware | Collect confidential information | Anti-virus and Malware detection system |
| XSS attack | Disclosure of sensitive information | Use of authorized apps |

### 2.4.3.1 Trojans

By using various mobile applications attacker can gain control over the device. Such kind of applications appear to perform some useful functionality in the foreground but perform malicious actions in background. Attackers can use Trojan to collect private information or for installing malicious applications like bots, or worms in the device. For example, a fake banking application can collect login information from the user. These applications mainly spread through third party application store [52]. Trojan can create a backdoor and spy on the device, turning the device into a zombie, or can use it to send costly messages.

**2.4.3.2 Worms**

Worm is a malicious self-replicating application that can spread into uninfected systems by itself without human intervention. For propagation, worm relies on vulnerabilities of networking protocols. Due to its replicating nature and capability to spread itself over the network, it damages and affects the security of the device and consumes the bandwidth of the network. With the introduction of Cabir, malware can be ported to the mobile devices. Cabir [53] worm attacks Symbian S60 devices, and it spreads through Bluetooth. Worm is observed only when its replicated instances consume system resources making the device slower.

**2.4.3.3 Rootkits**

It is a malicious application that runs under the privileged mode. Rootkit works in such a way that the user is not aware about the fact that operating system has been compromised. Rootkit itself is not harmful but is used by attackers to hide worms, bots and malware. It is not used to get access to the device but to hide malware efficiently [54]. During installing a rootkit, attacker has to get access to root account either using social engineering or by cracking the password. Since rootkits are initiated even before the operating system of the device, these are difficult to detect and remove [55].

**2.4.3.4 Mobile Ransomware**

It is a malware that locks the device of the user preventing him from accessing the data. Ransomware encrypts the data of infected system and decrypts the data only when ransom is paid to attacker. Ransomware attacks both computer as well as mobile phones. It locks the phone by changing the PIN number and then ask for ransom to unlock it. Ransomware can be of two types- crypto ransomware and locker ransomware. Crypto ransomware encrypts the file and data while locker ransomware locks the devices so that the owner of the device cannot access the data [56]. Table 2.3 shows the top 10 countries attacked by Ransomware [56].

**2.4.3.5 Botnets**

Botnet is a network of compromised computers called "Zombies", in which a malware called bot is installed in the device by the attacker. Main goal of this strategy is to utilize the computational power of the compromised machines to commit other activities. These compromised machines after installation of the malware into them are controlled by the attacker from remote locations. These devices are used to send a large number of spam emails, information theft, and can also be used to launch DDoS attacks. Most of the bots are developed by the attackers for financial gains [52].

Table 2.3 Top 10 countries attacked by Ransomware

| Rank | Country |
| --- | --- |
| 1 | USA |
| 2 | Japan |
| 3 | UK |
| 4 | Italy |
| 5 | Germany |
| 6 | Russia |
| 7 | Canada |
| 8 | Australia |
| 9 | India |
| 10 | Netherlands |

**2.4.3.6 Key Loggers and Screen Loggers**

Key loggers and screen loggers when installed in the device can cause a serious threat to it. They are distributed as malware and are not detected by anti-viral software. Key logger records the key pressed by the user with a keyboard of the compromised device and sends this recorded data to the attacker without user's knowledge. Screen logger records the screen and the mouse movement, thus making use of virtual keyboard unsafe.

**2.4.3.7 Spyware**

Attacker uses online available spyware to take over the mobile phone, with which they can control SMS, emails, listen to phone calls, and track victim's position using GPS. Concealed channels present in the smartphone are used by spyware for sending information to the attacker. When an application needs to send data to the outside world for legitimate purpose, the permission settings of the smartphones are not strict enough to prevent misuse of such authorization for any other purpose [57]. Zitmo spyware is one of the most dangerous spyware.

**2.4.4 Phishing through Online Social Networks**

For both professional as well as personal communication, social networking sites are used by millions of people around the world and have become an important part of the Internet. Social networking sites allow users to interact and share their ideas with other users. As large number of people are the part of these networking sites, it is a new ground of attacks for the attackers. Attackers exploit user's trust on social networking sites for their own benefits [58]. Some examples of social networking sites are Facebook, Twitter and MySpace. In 2016, phishing attacks on social media have increased by 150% [48]. Attackers are using social networking sites to initiate phishing attacks due to the popularity of these sites, and it is somewhat easy for an attacker to masquerade as someone else on these sites. 24% of the users click on fake social media connecting requests and half of these users even share their credentials [35]. The study shows that it is easy to trick the people on social networking sites. Also, Internet users are four times more likely to become a victim especially if they are asked by the person who claims to be their known. Various methods with which attackers can deceive users on social networking sites are discussed below.

**2.4.4.1 Impersonation**

Users like to follow the famous personalities on social networking sites and join groups of their interest. There is no procedure that verifies whether a virtual profile is actual or not [58]. The attacker makes use of this and pretends to be a famous personality and posts some malicious links regarding sale or offers which if opened, ask for personal details or end up downloading a malware [59].

**2.4.4.2 Posting Malicious Links**

Attackers use malicious links to redirect the user to some external malicious website which is under the control of the attacker. The links may be posted by dummy accounts. When a malicious URL is posted by the attacker, nearly 90% of the clicks occur within 24 hours after they are posted [48]. The redirection can be accomplished by social engineering where link appears to be a promising link. It is difficult for social media provider to block social engineering attacks as it is difficult to identify if it is legitimate or not. The redirected website may contain misleading information like malicious application, fake login page, or advertising fake products [58].

**2.4.4.3 Fake Profiles**

Attacker may send friend request to the users claiming to be their old friend. Once added to the friend list, attacker can access private information which the user shares with his friends, family and colleagues. For more information attacker may inbox user asking for phone number or email [60].

**2.4.4.4 Fake Communities**

Attacker may create a fake group with the name of well-known organization and add some members to the group who are already the part of that organization but are also with the attacker to carry out the scam. They send group request to other members of the organization who after seeing that their colleagues are also the members of the group, join the group. Attacker then obtains the secret information from their discussions and use it for his personal gain.

**2.4.5 Phishing through Content Injection**

In this, phisher modifies part of the content of the reliable website to deceive the user and take him out of the legitimate webpage where he is asked for the personal information. For instance, an attacker may inject malicious code to record user's information and deliver it to the attacker's server.

**2.4.5.1 Cross Site Scripting (XSS)**

XSS attack is an application layer web attack that targets vulnerable scripts embedded in the webpages that are executed on client side. JavaScript is used to deliver the malicious content to the users and alters the client-side scripts of the web application so that the script executes as per the attacker. XSS attacks have severe effects and results in compromise of user's account, modification of content of webpage, and revealing the credentials or session cookies [61].

**2.4.6 Phishing through Wireless Medium**

There are various kinds of wireless attacks that target the sensitive information of the user. Attackers spy on the data transmitted through the wireless medium to gain sensitive information of the user. These attacks can manipulate the hardware identification of the devices in order to spy on the user. Different security challenges in the wireless environment are discussed in [38]. Wireless attacks on mobile devices can be launched through Wi-Fi or Bluetooth.

**2.4.6.1 Wi-Fi**

Wi-Fi has become a vital part of our generation and as a result, hotspot for the attacker. Generally users do not authenticate the access point to which they are connecting and it is not difficult for an attacker to set up a bogus access point with Service Set Identifier (SSID) which looks similar to the legitimate one. Attacker can interrupt communication between mobile phone and the Wi-Fi hotspot [62]. Hotspot architecture is vulnerable, and it does not perform any encryption to protect the data being transferred. When the user connects to this type of hotspot for the first time, the connection between the two is not secure, and the attacker can interrupt and control the traffic, thereby hijacking the session.

**2.4.6.2 Bluetooth**

Bluetooth is a technology standard with which data can be shared over wireless links among the devices which are present within a short range. Devices having Bluetooth have some flaws due to which others can connect to the device without user's permission [62]. When the two mobile devices are in the range of each other, attacker's mobile device can send malicious data to the victim's device by establishing a Bluetooth connection using default Bluetooth passwords [52]. Once attacker gains access to the device via Bluetooth, contacts, messages, and files can be accessed by him.

**2.4.7 Phishing through Technical Subterfuge**

It is a method of tricking users in order to get some information that can be used by attackers for personal benefits [63]. Some of the mobile phishing attacks based on technical subterfuge are discussed below.

**2.4.7.1 Host File Poisoning**

A host file is the one which contains domain names and their corresponding IP addresses. When a client requests for a URL, it is first converted into IP address before transmitting it over the Internet. Host file poisoning involves altering the entries of the website in host file. This technique is used by attackers to perform phishing attack so that instead of redirecting the user to legitimate website, he is redirected to a fraud website where the user is asked for the personal information [64].

**2.4.7.2 Phishing through Search Engines**

Phishing attack also makes use of search engine that redirects the user to online shopping websites that offer services or products at low price. When the user tries to buy the product, the website collects the credit card details. Here the search engine is legitimate, but the website is fake, which is created by the attacker to steal personal information. The search engine may show discount offers, job offers, etc. to lure the victim [65].

**2.4.7.3 Using Compromised Web Servers**

Attacker looks for the vulnerable web server, and then compromises these web servers. Password protected backdoors are installed and attacker gains access to the server via encrypted backdoor. Pre-built phishing websites are downloaded and with the help of mass email tools, fake websites are advertised [66]. A study by Tyler Moore et al. [67] states that 76% of phishing website are hosted on compromised web servers.

**2.4.7.4 Man-in-the-Middle (MITM) Attack**

In this attack, attacker sits between the victim and the legitimate website. The data submitted to the legitimate website is received by the attacker which can be credit card details or any personal information. The attacker continues to pass the data to the legitimate website so that user's transaction is not affected. Secure Socket Layer (SSL) traffic is not vulnerable to MITM attack [68], but a malware based attack can modify the system configuration for installing a trusted certificate authority through which attacker can create his own certificate.

## 2.4.7.5 Secure Socket Layer (SSL) based Attacks

Phishing attack is carried out through untrusted websites. Though phishing website and original website look similar, but difference is that phishing websites do not have SSL certificate. SSL certificate is used by website operator to ensure data is transmitted over secure channels between browser and server. Phishing websites do not use SSL certificate based communication. Moreover after getting the credentials, phishing website may redirect user to original website having SSL certificate to fool users [69].

## 2.4.7.6 Session Hijacking

Session hijacking is exploiting valid computer session of the user. In this attack, attacker tries to obtain session id of the user in order to hijack the user's account. When a user submits his credentials, application server tries to authenticate him on the basis of the cookie values which consists of session id (SID). Hence, if the attacker gets SID of an active user, he can use it and login to the account, and gets access to details available in the account [70]. Cross-site scripting attack can also be used to hijack the session of the user.

## 2.4.7.7 Domain Name Server (DNS) Poisoning

It is a type of attack in which an attacker takes advantage of vulnerabilities present in domain name server to divert the incoming legitimate traffic towards the fake websites [71]. Whenever client browser requests for a domain name, the request is sent to the DNS server for getting corresponding Internet Protocol (IP) address. Attacker can set up a fake DNS server or alter the existing DNS table, changing the IP address corresponding to the domain name. Once the DNS is poisoned or fake entries are created, users are redirected to spoofed webpages [72]. Figure 2.10 shows process of phishing through DNS poisoning.

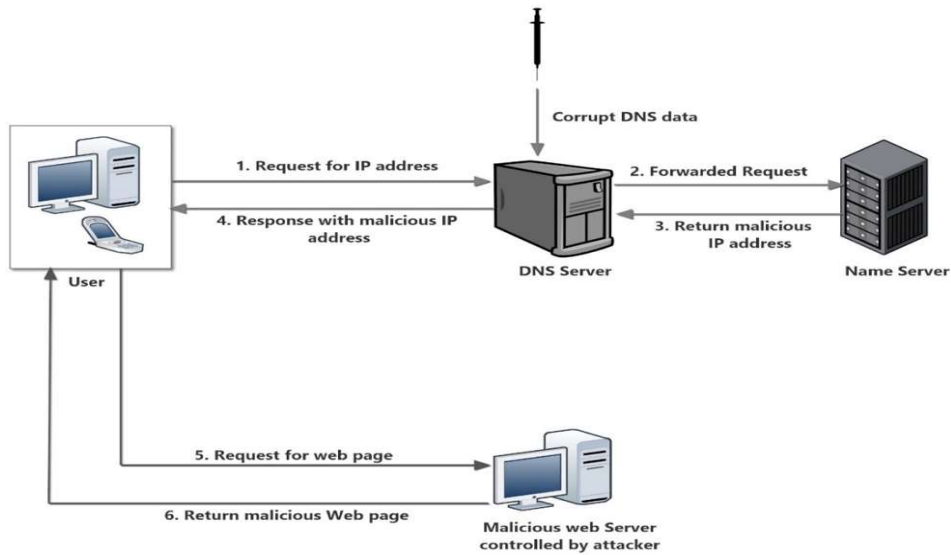

Figure 2.10: Phishing through DNS poisoning

## 2.5 Taxonomy of Mobile Phishing Defence Mechanisms

In this section, various approaches for detecting and defending against phishing attacks on mobile phones are discussed. Anti-phishing solutions for various mobile phishing attacks are presented in Table 2.7. Comparative analysis of various mobile phishing defence techniques is shown in Table 2.8.

### 2.5.1 User Education

Phishing is one of the most significant problems faced by Internet users. User education is important in order to create awareness among the users about phishing attacks. Phishing is a social engineering attack and targets users instead of devices. Hence, educating the user is important to avoid phishing attack. Education based approaches include showing warnings, and online training through games.

#### 2.5.1.1 Active and Passive Warnings

User interface shows warning depending upon the action triggered as deployed by many web browsers. The warning can be a passive warning that only shows the warning and relies on the users to perform certain action while active warning does not rely on the user to perform some action and block the content itself. Users do not pay attention to the warnings. Studies have shown that passive warnings are less effective as compared to active warnings.

### 2.5.1.2 Training through Mobile Games

One of the important factor in avoiding phishing attack is to train users which leads to correct identification of phishing and legitimate instances. Various methods are there to train the users. To develop conceptual knowledge about phishing attacks, various mobile games are being developed to educate the users. With the increase in use of Internet technology, the risk of mobile device users falling victim to phishing attacks have also increased. So mobile games can be used to train the users which further helps in reducing the phishing threat. Asanka et al. [73] designed one such game. The game is about educating the users about phishing emails and phishing URLs so that the user is able to differentiate between phishing and legitimate emails and URLs. The prototype of the game was implemented on Google App Inventor Emulator. In another approach, Asanka el at. [74] developed a game by identifying the elements that are needed to be addressed to avoid phishing attacks for educating users. In addition, Asanka et al. [75] designed a gaming approach by combining conceptual and procedural knowledge to educate users. The approach integrates "self-efficiency" to the anti-phishing educational game in order to enhance user's behavior to avoid phishing attacks. Other game based approaches are presented in [76-79].

### 2.5.2 Detection of Smishing and Spam SMSes

Smishing messages consist of a text message and a URL which when opened perform malicious activity. Attackers use social engineering approach to target victims and users are easily attacked by it. For detecting smishing or spam messages, different classifiers that make use of effective feature set are used. Table 2.4 shows the comparative analysis of various smishing detection approaches. Various approaches for detecting smishing and spam messages are discussed below.

### 2.5.2.1 S-Detector

Joo et al. [7] proposed a security model "S Detector" for detecting and blocking smishing messages. It is a content based analysis approach. Naïve Bayesian Classifier is used to differentiate between smishing and normal messages by extracting the words most often used in smishing messages. S-Detector consist of four components - SMS monitor, SMS determinant, SMS analyzer, and Database. S-Detector takes the following steps to distinguish normal messages from Smishing messages.

1) When a text message is received, SMS monitor records the logs and timestamps of the communicated SMS message.
2) It is checked if the telephone number is already registered in blacklist database.
3) It is determined if the text message contains a URL. If yes, accesses that URL.
4) It is checked if an APK file is downloaded on access to the URL. If an APK file is downloaded, it is regarded as smishing message and is blocked, else content of the message is analysed.
5) Pre-processing is done to separate the strings from the text message and morpheme unit are extracted. Then a weight value is assigned to each word using Naïve Bayes algorithm.
6) If weight is more than or same as threshold, the message is labelled as a Smishing message and is blocked. Otherwise, it is categorised as a normal message.

**2.5.2.2 SMSAssassin**

Yadav et al. [80] proposed a mobile spam messages filtering application "SMSAssassin" based on Bayesian learning. Support Vector Machine (SVM) is used along with Bayesian learning in order to achieve higher accuracy. Spam SMS consist of patterns and keywords that are changed frequently. Crowd-sourcing is used to keep the list of patterns updated. During the training stage, the occurrence of each word in spam and ham messages is computed to determine whether a word belongs to ham or spam. After training, the spaminess probability of SMS is calculated, and if it is above a certain threshold, then it is regarded as spam message. To keep track of spam keywords, SMSAssassin uses GlobalSpamKeywords at the server and SpamKeywordsFreq list in mobile phones. The mobile application also maintains a UserPreferencesList under which user can mention ham/spam keywords according to his choice or preferences. Users having SMSAssassin application in their mobile phones can share reported spam list. Authors collected a total of 4318 SMSes using crowdsourcing. Bayesian learning technique gives 97% classification accuracy in ham SMSes, 72.5% classification accuracy in spam. Table 2.5 shows the list of effective features for detecting spam messages [82].

Table 2.4 Comparison of various smishing detection approaches

| Security features | Joo et al. [7] | Yadav et al. [80] | Lee et al. [87] | Smishing defender [85] |
|---|---|---|---|---|
| Check for presence of URL | ✓ | x | ✓ | ✓ |
| Check for APK download | ✓ | x | ✓ | ✓ |
| Check for login page | x | x | x | x |
| Check for sender's mobile number | ✓ | ✓ | ✓ | ✓ |
| Check for self-answering messages | x | x | x | x |
| Text normalization | x | x | x | x |
| Content based analysis | ✓ | ✓ | ✓ | ✓ |

**2.5.2.3 Dendritic Cell Algorithm (DCA) based Approach**

Sayed et al. [81] proposed a technique for filtering multimodal textual messages including emails and short messages. Inspired from the human immune system and hybrid machine learning methodologies, the author proposed a method for information fusion. Various features obtained from the received messages were analyzed with the help of machine learning algorithm. They developed a framework based on DCA for mobile spam filtering by fusing output from machine learning algorithms.

**2.5.2.4 Text Normalization and Semantic Indexing based Approach**

Almeida et al. [83] proposed a mechanism that normalizes and expands the short and noisy text messages. Semantic and lexicographic dictionaries are used for this purpose. The text is processed in three stages- text normalization, concept generation, word sense dis-ambiguity. Text normalization normalizes and translates each term into its canonical form and uses two dictionaries - first is English dictionary and second is lingo dictionary. Concept generation is used to obtain every meaning or concept related to a particular term. Word sense dis-ambiguity is used to find the most relevant concept or meaning according to the context of the message. Concept generation and word sense dis-ambiguity uses LDB BabelNet Repository. Authors concluded that with the help of text processing, classification performance can be enhanced. The system improves the quality of the attributes obtained, which in turn improves the classification accuracy.

Table 2.5 List of features for detecting spam messages

| S. No. | Feature | Description |
|---|---|---|
| 1 | No. of characters in message | Length of message on basis of number of characters |
| 2 | No. of words in message | Length of message on basis of number of words |
| 3 | Frequency of word "money" | Number of times word "money" appears in message |
| 4 | Frequency of symbol "money" | Number of times symbol "money" appears in message |
| 5 | Words in capital letters | Count the number of words that appears in capital letters |
| 6 | Number of special character | Number of special character appears in message |
| 7 | Number of emoticon symbol | Number of times emotional symbol appears in message, generally used by legitimate user |
| 8 | Presence of links | Check for the presence of link in message, mostly used by attackers |
| 9 | Presence of phone number | Check for the presence of Phone number in message, generally used by attackers |
| 10 | Average number of words | Ratio of number of words to number of characters in message |
| 11 | Number of sentence | Total number of sentence present in message, sentence tokenizer can be used |

### 2.5.2.5 Spam detection using Content of Text

Karami et al. [84] proposed a content based approach which instead of depending on individual word, uses a semantic group of words as features. Linguistic Inquiry and Word Count (LIWC) and SMS Specific (SMSS) features are the two semantic categories of features used by the researchers that helps to reduce the feature set, in turn improving the efficiency of the approach. There are two phases in the system – feature extraction and classification. Machine learning algorithm is used for classification. Accuracy of the system lies from 92% to 98%.

### 2.5.2.6 Smishing Defender

An application "Smishing defender" was developed by Hauri Inc. that detects and blocks phishing SMS messages in Android smartphones. The application monitors the text messages received and notifies the user on the reception of smishing message in real time. The application also provides a feature with which suspicious messages can be sent to Hauri for further analysis of the messages [85].

### 2.5.2.7 MDLText based Approach

Increasing volume of data in smartphone devices requires efficient and effective text classification methods. Silva et al. [86] developed "MDLText" which is an efficient, scalable, fast, and lightweight multinomial text classifier based on the Minimum Description Length principle. MDLText is robust, learns faster and avoids over-fitting problem. Due to incremental learning, the scheme can be used in online as well as dynamic scenarios. Even with large volume of data, MDLText has lower computational cost.

### 2.5.2.8 Detecting Smishing Attack in Cloud Computing Environment

Lee et al. [87] proposed a technique to detect smishing messages using cloud virtual environment. The proposed technique checks for source of the message, content and location of the server and takes decision accordingly. Smishing detection probability is increased by using program interface analysis and filtering so as to minimize incorrect detection. On receiving a message, the user can compute the risk of the message in virtual environment and processing is also done there. When the process is completed, the screenshot and the report is sent to the user. Based on the report, user can determine if the message is smishing or not which in turn reduces the incomplete and false detection.

### 2.5.2.9 Feature based Framework for SMS Spam Filtering

Uysal et al. [88] proposed a framework for SMS spam filtering. To find various features of SMS, it uses two feature selection methods that are based on chi-square metrics and information gain (IG). Features are fed to the Bayesian classifier to classify the SMS as ham or spam. The scheme was designed for android mobile phone users and evaluated on large set of SMSes including legitimate and spam messages and output shows that system gives accurate results in detecting both ham as well as spam messages.

### 2.5.3 Detection of Phishing Webpages

Website phishing attack targets the users instead of systems and it is comparatively easy to carry out these attacks as creating an exact replica of a website is not a difficult task. Phishing websites are difficult to detect due to high level of similarity they possess with legitimate websites. So some techniques are required for detecting website phishing attacks. Some of the existing methods are discussed below.

**2.5.3.1 MobiFish**

Wu et al. [47, 89] proposed a lightweight anti-phishing scheme "MobiFish" for mobile platforms that protects users from phishing attack on mobile applications, webpages, and persistent accounts. The fundamental idea behind this scheme is that if there is any dissimilarity between the identity an instance actually is and the identity it claims to be then it leads to a phishing problem. The claimed identity is obtained from the screen presented to the user. Screenshot of the login screen is taken and Optical character recognition (OCR) tool is used to extract text from the screenshot. The actual identity of the mobile webpage is obtained from the Second Level Domain (SLD) name. If the second level domain is not present in text obtained from the screenshot, then it is an unsafe page and MobiFish will warn the user else webpage is a safe page. The authors implemented MobiFish on a Google Nexus 4 smartphone which used android 4.2 Operating System. MobiFish performance was evaluated on 100 phishing URLs and their corresponding legitimate URLs as well as on phishing applications. The experimental results validated that the MobiFish method can effectively fight against the phishing attacks.

**2.5.3.2 Feature based Detection**

Tripathi et al. [90] proposed a technique to detect malicious mobile webpages. This technique uses both static as well as the dynamic features to determine legitimacy of the webpage. The system uses OCR tool to convert screenshot of the webpage into text. When the user enters the URL, system analyzes HTML code, JavaScript content, and checks for cross site scripting. Then it is checked if the second level domain is present in the text obtained by converting the webpage screenshot into text. If the second level domain is present in the text, then the webpage is safe, else warning message is displayed to the user. This technique provides higher accuracy and better classification. The browser extension is built to achieve real time feedback.

**2.5.3.3 MP-Shield**

Bottazzi et al. [91] proposed MP-Shield, a framework for detecting phishing attacks on mobile webpages. It is an android application that is implemented on Transmission Control Protocol/

Internet Protocol (TCP/IP) stack as proxy service. Its aim is to inspect IP packets originating from and directed to the mobile applications by extracting HTTP get request from the packet. Virtual private network (VPN) service is used to inspect the packets and watchdog engine is used to check if the URL is blacklisted or not. If URL is blacklisted then a warning is shown to user, else classification engine is used to classify URL as legitimate or suspicious. MP-Shield provides high level of security without causing disturbance to the user.

**2.5.3.4 kAYO**

Amrutkar et al. [10] designed and implemented kAYO, a browser extension that differentiates between benign and malicious mobile webpages. It is based on the idea that mobile webpages are different from their corresponding desktop webpages. kAYO uses static features of webpages to identify malicious mobile webpages. They used 44 mobile specific features of webpages and out of these 11 are newly addressed features. These features are derived from URL, JavaScript content, HTML and mobile specific capabilities. The feature set includes 12 URL, 10 JavaScript, 14 HTML and 8 mobile specific features. The list of effective features of mobile webpage are discussed in Table 2.6 [10]. kAYO browser extension works as follows -

1) The user enters the URL in the extension toolbar.
2) Extension toolbar sends URL over Hyper Text Transfer Protocol Secure (HTTPS) to kAYO's backend server.
3) The URL is crawled by the server, and static features are extracted.
4) The feature set is given to kAYO's trained model, and the model then classifies webpage as phishing or legitimate and decision is sent to user's browser.
5) If URL is benign then the extension will render webpage in the browser automatically, else shows a warning message.

Authors collected over 350,000 benign and malicious mobile webpages and identified new static features from these webpages that can distinguish benign webpages from malicious webpages. This data set is used to train kAYO's model. kAYO gives 90% classification accuracy, 89% true positive rate and 8% false positive rate. kAYO browser extension is designed for Firefox Mobile browser. Technique is fast and reliable. Due to efficient feature set, this technique is able to detect new mobile threats that Google safe browsing and VirusTotal cannot detect.

**2.5.3.5 URL based Detection**

Chorghe et al. [92] proposed a technique for detecting phishing attacks on android mobile devices. This technique is also capable of identifying Zero-day phishing attacks. It consists of five components - extraction of URL, static analysis of URL, foot-printing of webpage, URL based heuristics, and SVM classifier. Blacklist, static analysis, and machine learning algorithms are used to get precise results. It was implemented on Moto G3 android phones having 6.0.1 Android version and system achieved 92% accuracy.

**Table 2.6** List of effective features for detecting webpage phishing

| Type | Features |
| --- | --- |
| JavaScript based features | Presence of JavaScript, internal JavaScript, embedded JavaScript, external JavaScript, noscript; number of JavaScript, external JavaScript, internal JavaScript, embedded JavaScript, noscript |
| HTML based features | Presence of images, internal and external links; number of images, external links, internal links, Number of HTTPOnly cookies, number of cookies from header, presence of iframes and redirections, whether webpage served over Secure Socket Layer, number of iframes and redirects, percentage of white spaces in HTML content |
| URL based features | Number of deceptive terms in URL like bank and login, length of URL number of digits, forward slashes, question marks, hyphens, underscores, dots, number of equal signs, subdomains, ampersand, subdomains with two letter, semicolons, presence of subdomain, percentage of digits in hostname |
| Mobile specific features | Number of API calls to sms:, tel:, smsto:, mmsto:, mms:, geolocation;, number of ipa, number of APK |

**2.5.3.6 Phishing Blacklist**

Most of the popular anti-phishing techniques make use of blacklist approach. Blacklist is a list of suspicious IP addresses, URLs, and keywords. It determines if the URL is fraud or not. Fraud URL means it is used by attackers for stealing user's information. Blacklist needs to be frequently updated as it can only detect phishing URLs and IP addresses listed in it [93]. Blacklist is a valuable source used by anti-phishing toolbars to warn users and deny access to fraud phishing websites.

Microsoft has used blacklist based anti-phishing solution with Internet Explorer 7. Blacklisting approach is currently used by Chrome, Google search, and Gmail. Limitation of the blacklist is that it does not detect zero day phishing attacks and can identify only 20% of these attacks. The approach works well as long as the list is regularly updated. This method has high accuracy with a less false positive rate.

Table 2.7 Anti-phishing solution for various Phishing techniques

| Phishing Techniques | Anti-Phishing Solution |
|---|---|
| Smishing | Dynamic models and framework |
|  | Blacklist |
| Vishing | User training |
|  | Blacklist |
| Phishing websites | Password management tools |
|  | Trusted path ensured browser |
|  | Client server authentication |
|  | Browser extension |
|  | Pattern matching |
|  | Blacklist |
| Phishing and spam emails | Anti-spam filters |
|  | Client server authentication |
|  | User training |
| Phishing applications | Personalized security indicators |
|  | Permission based analysis |
| Malware | Anti-malicious programs |

### 2.5.3.7 Whitelist

Whitelist contains list of legitimate URLs, and it is opposite of blacklist. Whitelist proposed by Cao et al. [94] maintains a list of legitimate LUIs and confirms the legitimacy of trustworthy pages that are already present in the whitelist. It prevents users from giving credentials to unrecognized or unauthorized websites. When the user enters their credentials in the trusted LUIs present in whitelist, system will not show any warning. But if there is an attempt to submit sensitive data to LUI that is not in the whitelist, then user's browsing will be stopped and user will be warned about the forgery [95]. Maintaining a whitelist is easy as compared to blacklist because legitimate website hardly change their URLs [96]. Using the similar concept, Han et al. [97] proposed an anti-phishing method for smartphone devices. The main concern is that user is not able to differentiate fake LUI from genuine LUI. So, Han et al. proposed a method in which smart devices store the feature information of LUI in advance. Before the user enters authentication information [98] in LUI, a browser plug-in verifies the LUI based on pre-stored LUI information. If LUI passes the verification of smart device, then the device allows to fill the user id and password field. This method significantly improves the security of Login information.

### 2.5.4 Detection of Malicious Mobile Applications

Phishing attacks via mobile applications is another major problem faced by mobile device users. Once the malicious application enters the device, they control device of the user and collect personal information and send it to the attacker. When applications are installed, they request for many irrelevant permissions that have nothing to do with the working of application. Various approaches to detect malicious applications and methodologies to prevent data leakage from the mobile applications are discussed below.

### 2.5.4.1 VeriUI

Liu et al. [99] proposed a system "VeriUI" based on attested login, which is a password protection mechanism for mobile devices. It provides a secure and hardware isolated surroundings to input password so as to prevent phishing attacks that occur through mobile applications. When the user logs in through VeriUI device, this scheme supplements credentials with software and hardware information of a device and some contextual meta-data. VerUI assures that credentials handling is separated from the rest of application and are processed in a secure environment.

### 2.5.4.2 StopBankun

There are some malicious applications that aim to steal user's credentials and attack existing mobile applications by replacing them with the modified applications. Kim et al. [100] proposed a method StopBankun to prevent replacement of banking applications with the malicious applications. Whenever a third party application attempts to remove a legitimate application, PackageInstaller is launched that shows very limited information due to which unaware users allow it to perform actions and the application is removed from the device. The objective of StopBankun is to make PackageInstaller show sufficient information so that the user is able to notice any illegal attempt. Process of removing an application from mobile phone is shown in Figure 2.11.

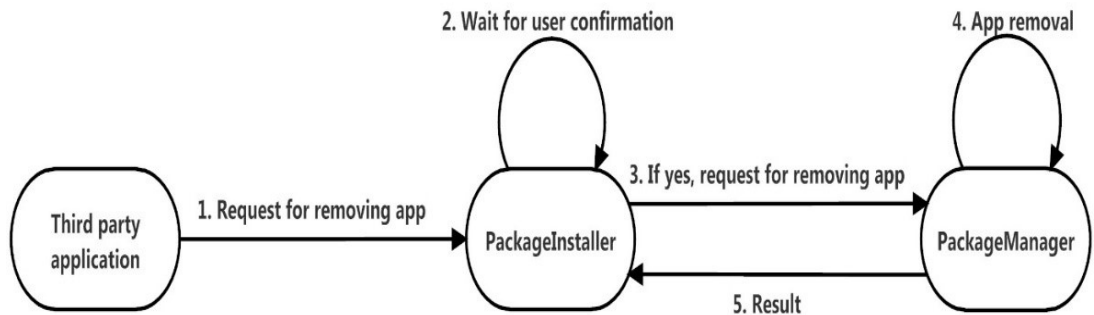

Figure 2.11: Application removal procedure

### 2.5.4.3 Confidential Data Leakage Prevention

Smartphones store huge amount of personal information, such as messages, contacts, etc. and this information is very sensitive. Malicious applications try to access this information via permissions. Canbay et al. [101] proposed a system for Android devices to prevent confidential data leakage. It uses J48 classification algorithm to detect applications that leak sensitive information. K-Means clustering algorithm is used to evaluate mobile applications downloaded from google play store to check if they have any resemblance to the malicious mobile applications. The system effectively detects malicious and benign applications with 98.6% accuracy.

### 2.5.4.4 Smartgen

URL plays an important role in the detection of phishing attacks but URLs are hidden in mobile applications, unlike desktops where URLs are directly visible. So a mechanism is required with which we can see the URLs in mobile applications. Zuo et al. [102] developed a tool SMARTGEN that automatically sends a request message to the server and retrieves the URL for the mobile application. This URL is then submitted to a service named "VirusTotal" that detects harmful URLs for security analysis.

### 2.5.4.5 Andromaly

Shabtai et al. [103] proposed a light-weight approach for detecting host based malware. The technique observes various features and events of the mobile device and use machine learning detectors to classify data collected as phishing or legitimate. The results have shown that the proposed technique is effective in detecting malicious software on mobile devices. With the help of Andromaly, suspicious behavior of an application can be reported to the Android market.

**2.5.4.6 AppsPlayground**

Rastogi et al. [104] proposed a framework AppsPlayground for detecting malicious functioning and privacy leakage from the Android applications. AppsPlayground incorporates many components for detection of malicious applications. A wide range of detection techniques are used like Taint Droid, kernel level system monitoring. AppsPlayground analyzes for both malware as well as Grayware. This tool effectively evaluates android applications at large scale.

**2.5.4.7 TrueWalletM**

In order to protect login credentials that are used to access online services over the mobile devices, Bugiel et al. [105] proposed a secure wallet based system "TruWalletM". This approach uses Trusted Execution Environment and hardware security features present in mobile phones. The method works well with legacy software and standard authentication methods without imposing any performance overhead. Communication process is slightly slowed down. The authors have addressed the issue of run time compromise of interface which results into credentials disclosure. It is a password manager and authentication agent which protects login credentials without requiring the user to trust the operating system.

**2.5.4.8 RiskRanker**

Grace et al. [106] proposed a proactive approach to detect zero-day android malware. This approach analyze android applications according to some set of rules that are specifically designed to detect the malicious behaviour of the application and this behaviour is called as potential risks associated with untrusted applications. The risk is divided into three categories - high-risk, medium risk, and low risk. RiskRanker after analyzing dangerous behavior, generates a prioritized list of applications which can be further examined.

## 2.5.5 Quick Response (QR) Code based Techniques

QR-code is a two-factor authentication scheme secure under untrusted local systems and servers. It is used to trace the products and store massive data about them. QR code [107] is used in many industries, such as an advertisement, sale wrapping, and business cards.

### 2.5.5.1 Single Sign-On based on QR-code

Mukhopadhyay et al. [108] proposed a secure single sign-on method that uses mobile QR-code based one-time password scheme. Single sign-on allows a user to login once and access multiple services. Single Sign-On method do not protect users from real-time attacks but in the proposed scheme, the authors gave an anti-phishing Single Sign-On solution which is effective against phishing attacks.

### 2.5.5.2 Authentication Scheme using QR-Code

To protect the personal information from phishing attacks on mobile devices Choi et al. [109] proposed a Single-Sign-On authentication scheme based on QR-code. This scheme has addressed the limitation of Single-sign-on which allows the user to access multiple applications with single username and password. In the proposed approach, server generates a random key which is used for secure communication. This scheme works in three phases. First is login request phase, second is QR-code generation phase, and third is verification phase. The scheme encrypts the information due to which attacker cannot obtain the information even if the information is exposed to them.

## 2.5.6 Personalized Security Indicators based Techniques

Personalized security indicators are the visual indicators shared between the user and the application. With the help of these indicators, users can identify genuine applications even if malicious applications are present. When the user starts an application for the first time, he is asked to choose an indicator for each application. Marforio et al. [49,110] proposed a scheme using personalized security indicators for detection of phishing attacks on mobile applications. With the help of these security indicators, a user can distinguish genuine applications from malicious applications. Various online services are also using personalized security indicators.

**Table 2.8** Comparisons of various existing solutions for mobile phishing attacks

| Approach | Techniques | Specific for | Advantages |
|---|---|---|---|
| S-Detector [7] | Naive Bayes classifier | SMS | Approach is able to detect and block smishing messages with high accuracy rate. |
| Kayo [10] | Machine Learning | Mobile Webpages | 90% classification accuracy, 89% true positive rate, 8% false positive rate. |
| MobiFish [47] | Optical character recognition | Mobile webpage, application, and persistent account. | Web-Fish achieves 100% verification rate. |
| SMSAssassin [80] | Bayesian learning and sender blacklisting mechanism | SMS | 72.5% and 97% classification accuracy in spam and non-spam messages respectively. |
| Sayed et al. [81] | Dendritic cell algorithm | Emails and SMS | The dendritic cell algorithm improves recall and precision of spam and non-spam messages; accuracy approx. 100%. |
| Almeida et al. [83] | Text processing with lexicographic and semantic dictionaries | SMS | For the Wilcoxon Signed-Ranks Test, the null hypothesis is rejected with α=0.05 with a confidence level of 95%. |
| MDLText [86] | Minimum description length principle | SMS | Able to process high dimensional data at fast speed; Low computational cost. |
| Tripathi et al. [90] | Machine Learning | Mobile Webpages | High classification accuracy; browser extension for real-time feedback |
| MP-Shield [91] | Blacklist and data mining approach | Mobile Webpages | Ensure zero-hour protection; Protect android devices from phishing attack. |
| Chorghe et al. [92] | SVM classifier and URL based heuristics | Mobile Webpages | 92% accuracy and protects from zero-day phishing attacks. |
| VeriUI [99] | Augments user's credentials with hardware and software information | Mobile Application | Prevent phishing attacks through secure, hardware isolated environment for password input and transmission. |
| Canbay et al. [101] | J48 classification and K-Means clustering algorithm | Mobile Application | Approach can detect malign and benign applications with 98.6% accuracy. |

| Mukhopadhyay et al. [108] | Mobile QR code | Webpage authentication scheme | Protect against man in the middle attack and replay attacks. |
| Choi et al. [109] | Mobile QR code | Webpage authentication scheme | Data is encrypted so credentials are safe even if attacker obtain them; user can check if server is phishing or not. |

## 2.6 Research Issues and Challenges

Many solutions have been given by the researchers to detect and prevent mobile phishing attacks but still, there is no single solution that can detect or prevent all the attacks. Whenever researchers come up with any idea to fight against phishing attacks, attackers change their attacking strategy and find the weaknesses in the current solution. In this section, we have discussed some open research issues and challenges that are needed to be addressed.

- *Zero-day phishing attacks* – Foremost issue is to detect and prevent Zero-day phishing attacks. Zero-day attacks are those attacks that take advantage of vulnerabilities on the same day they are made public or in other terms that take advantage of publically known unpatched vulnerabilities. Zero-day attack may result into installation of malware, spyware, or unwanted access to user's personal information [111]. Currently, there is no solution available that can detect zero-day phishing attacks with high accuracy.
- *Determining appropriate threshold values* – Threshold is the level of similarity between two instances. Phishing sites look similar to their corresponding legitimate sites. Therefore, level of similarity between the two instances is calculated and compared with threshold value. So, to determine the appropriate threshold value is an important issue in order to get correct results. If the threshold value is greater than the appropriate value, then the false negative rate increases and if the threshold value is smaller than the appropriate value, then the false positive rate increases. A good anti-phishing solution is one having false positive and false negative rate as minimum as possible [26].
- *Language dependencies* – Language dependency is another issue. Different text languages are used over the Internet in websites, and mobile applications. There are some techniques, such as heuristics techniques that are language dependent as they use keywords and feature set. So language dependent techniques fail to work in the situations where a different language is used [26].

- *Selection of appropriate classifier* – Selection of appropriate classifier in machine learning based defending approaches is a challenging task. The good classifier takes minimum training time and gives high detection accuracy. For example, Support vector machine (SVM) is used to classify an instance as phishing or legitimate. But the problem with SVM is that it works for small datasets only. The naïve Bayesian classifier is another popular classifier, but it can be used only when the values of the various features are mutually independent [10].
- *Computational Requirements* – Smartphones have less computational and processing power. The phishing detection techniques developed for smartphone devices should have less computational requirements so that technique can work for these devices.
- *Real Time Detection* – If the technique is able to detect phishing attack and takes decision to either delete or block them in real time, then it will be very useful for the users. In such cases, only ham messages will be delivered to the users and users are not aware that they are receiving smishing messages.
- *Accuracy* – Accuracy is an important factor. The technique developed should give reasonable amount of accuracy with low false positive and false negative rate. Although blacklist and whitelist approaches have low false positive rates but these are inefficient in detecting zero-hour phishing attack.
- *Lack of awareness among users* – User awareness about phishing attacks is also an important issue [112]. Most of the non-technical users do not want to learn, and out of those who learn, most of them do not retain their knowledge for a long time. Some improvements should be made in the user interface. Active warnings should be given instead of passive warnings.

## 2.7 Chapter Summary

In this chapter, we provide details about the historical background of phishing attacks including evolution of phishing attacks and some recent statistics. We also discussed in detail the life cycle of mobile phishing attack. We analyzed various mobile phishing attacks and presented a taxonomy of the same. We also outline numerous techniques proposed by the researchers that detect and defend users from the mobile phishing attacks. Lastly, we discussed some research issues and challenges associated with mobile phishing attack that needs to be resolved.

# CHAPTER 3

# SMISHING SECURITY MODEL

SMS is one of the most widely used text based communication services. Availability of these text messages at low cost has attracted attackers to carry out phishing attack using SMS. In this chapter, we discuss our proposed text normalization based smishing security model in detail including its working algorithms. We have used machine learning classification algorithm to classify the message as smishing or ham message.

## 3.1 Proposed Solution

The objective of our proposed scheme is to classify the text message as smishing message or ham message but presence of abbreviations, short forms and slang terms in the messages makes it difficult to determine if a message is smishing message or not, which in turn reduces the classification accuracy of the scheme. To address this limitation, we have used a lingo dictionary to replace the abbreviations and short forms into their standard form. In addition, Naïve Bayesian Classifier is used to classify the message as ham or smishing message. The goal of our scheme is to detect smishing messages with high accuracy. The proposed smishing security model is shown in Figure 3.1. In this section, working of the proposed model is discussed. It consists of two phases – Preprocessing and Normalization phase, and Classification phase.

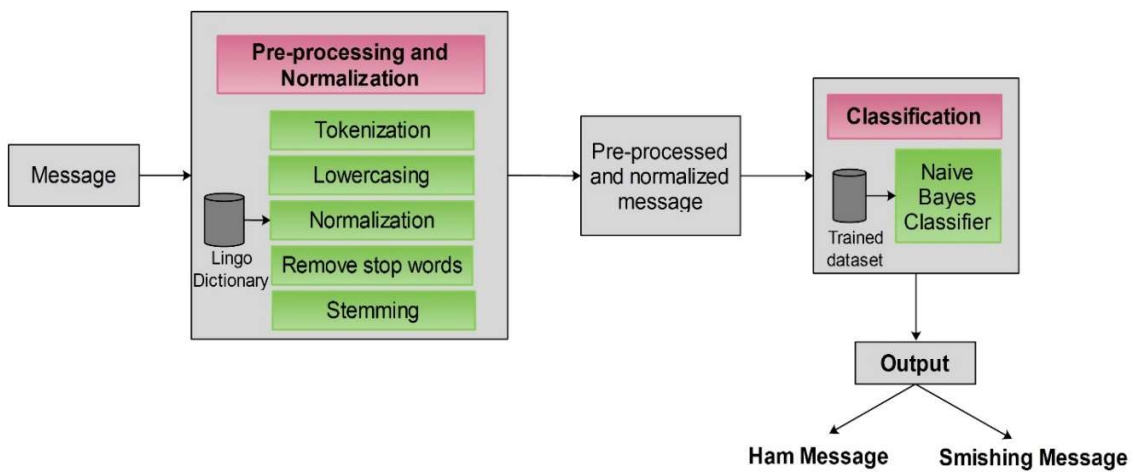

Figure 3.1: Proposed Smishing Security Model

## 3.1.1 Phase 1: Preprocessing and Normalization

When an SMS is received, the text message is first pre-processed and normalized. Pre-processing removes the unnecessary terms present in the message and normalization process replaces the obfuscated noisy text by its standard form. Pre-processing involves tokenization of strings, lowercasing of the text message, removal of stop words from the message and stemming. Additionally, in order to remove the short forms and abbreviations from the text message, the terms present in the text message are normalized to their standard form. We have used Lingo dictionary named NoSlang dictionary [113] for this purpose. Figure 3.2 shows the example of Pre-processing and Normalization phase. The output of this phase is pre-processed, normalized and clean text, which makes further processing of text easier for next phase. The complete procedure of Phase 1 is explained in Algorithm 1. The steps involved in this phase are discussed below –

- ***Step1: Tokenization -*** It is the process of splitting the text message into tokens. In this step, the sequence of characters is converted into sequence of tokens. Tokens are separated by whitespace and punctuation marks.
- ***Step 2: Lowercasing -*** In this step, the tokens obtained from the tokenization step are converted into lowercase tokens. This step is executed to make the text consistent.
- ***Step 3: Normalization -*** In this step, text message obtained from previous steps is normalized and replaced by its standard form. We have employed Lingo dictionary for this purpose, named NoSlang dictionary. This step works by looking for each token present in the text message in the lingo dictionary. If the word is present in the dictionary, it is replaced by its standard form and if the word is not present in the dictionary, the word is kept as original.
- ***Step 4: Removal of stop words -*** In this step, stop words present in the text message are removed. This step is performed to remove the unnecessary words present in the message as stop words do not contribute in the detection of smishing attacks. We have used the Natural Language Toolkit (NLTK) stop words list. This list consists of 153 stop words of English language that are commonly used articles, common words, pronouns and prepositions.
- ***Step 5: Stemming -*** In this step, the words are reduced to their base or root form. Stemming is done to avoid ambiguity. For example - bank, banking, and banks will be reduced to their root form – bank. It increases the accuracy of the scheme as all these three different words will be considered as one word.

```
Text message: Plz reply us with ur Bank Details
After tokenization: 'Plz', 'reply', 'us', 'with', 'ur', 'Bank', 'Details'
After lowercasing: 'plz', 'reply', 'us', 'with', 'ur', 'bank', 'details'
Normalized output: please reply us with your bank details
After removing stop words: please 'reply us bank details
After stemming: pleas repli us bank detail
Final preprocessed and normalized output: pleas repli us bank detail
```

Figure 3.2: Example of Preprocessing and Normalization phase

*Algorithm 1: Preprocessing and Normalization Algorithm*

**Input:** msg (message), dict (NoSlang dictionary), stop (stop words)

**Output:** n_msg (preprocessed and normalized message)

**begin**

    msg ← msg.tokenization

    msg ← msg.lowercase

    **for** each w in msg

        **if** w found in dict

            g ← g.append (standard form)

        **else**

            g ← g.append (w)

        **end**

    **end**

    msg ← g

    **for** each w in msg

        **if** w found in stop

            msg ← msg.remove(w)

        **end**

    **end**

    msg ← msg.stem

    n_msg ← msg

**end**

**3.1.2 Phase 2: Classification**

This phase involves classification of the message as smishing message or ham message. In this phase, the dataset is first pre-processed and normalized using the same process as discussed in phase 1. This is done to make the dataset consistent as those of text messages. Bayesian learning is used to train the dataset. In Bayesian learning, ham and smishing probability of each word is calculated and two separate files are maintained for training dataset - one for ham probability of each word and other file for smishing probability of each word. For example, words like bank, free, offer, call have higher smishing probability due to frequent occurrence of these words in smishing messages. On the other hand, these words have lower ham probability. Words like good morning, shop, college have higher ham probability as compared to smishing probability.

Naïve Bayesian Classifier is used to classify the message as ham message or smishing message on the basis of the trained dataset. This classifier is based on Bayes theorem with independence assumption among the features. It works better in complicated actual conditions. The pre-processed and normalized text message is given as an input to the Naïve Bayes classifier. After getting ham and smishing probability of each word of the message, ham and smishing probability of complete message is calculated using Naïve Bayes equation. The equation for Naïve Bayes Classifier is shown below.

$$p(C_k|x) = \frac{p(x|C_k)p(C_k)}{p(x)} \qquad (1)$$

Where x is an attribute value of the given data.
$p(C_k|x)$ is probability that data x belongs to specified class $C_k$ i.e. posterior probability of class $C_k$ given predictor x.
$p(x|C_k)$ is a likelihood i.e. probability of predictor x given class $C_k$.
$p(C_k)$ is prediction before the corresponding work occurs i.e. prior probability.

After this, smishing and ham probability of the message is compared, and if smishing probability of the message is greater than the ham probability then the message is categorized as smishing message, otherwise message is categorized as ham message. The complete procedure of Phase 2 is explained in Algorithm 2.

*Algorithm 2: Classification Algorithm*

**Input:** D (dataset), n_msg (preprocessed and normalized message)
**Output:** ham or smishing message
**Begin**

    n_Dataset ← apply algorithm 1 on D
    $D_{train}$ ← select and extract 90% of n_Dataset
    $D_{test}$ ← select remaining messages of n_Dataset
    **for** each message m in $D_{train}$
        **for** each word w in message m
            w_ham ← count no. of ham messages in which w appears
            w_smish ← count no. of smishing messages in which w appears
            w_ham_prob ← (w_ham) / total no. of ham messages
            w_smish_prob ← (w_smish) / total no. of smishing messages
        **end**
        ham_train ← ham probability of each word
        smish_train ← smish probability of each word
    **end**
    ham_prob_msg ← apply equation 1 on n_msg for k = ham
    smish_prob_msg ← apply equation 1 on n_msg for k = smish
    **if** (smish_prob_msg) > (ham_prob_msg)
        output ← smish message
    **else**
        output ← ham message
    **end**
    **return** output
**end**

## 3.2 Chapter Summary

In this chapter, we discussed our proposed smishing security model in detail. We preprocessed and normalized the text message to convert it into its standard form. Standardization of the text message will enhance the classification accuracy of the scheme. In addition, working algorithms are also discussed.

# CHAPTER 4

# EXPERIMENTS AND RESULTS

In this chapter, we discuss in detail the performance evaluation metrics used while formulating the proposed model. In addition, we discuss the dataset used in the model along with the results obtained from the experiments analysis. Furthermore, we present a comparative analysis of our proposed approach with some of the other related solutions.

## 4.1 Performance Evaluation Metrics

Identifying smishing messages is a binary classification problem where the objective is to identify smishing messages from a set of smishing and ham messages. The metrics is shown in Table 4.1 and is calculated as follows –

- ***True Positive (TP)*** – TP is number of smishing messages correctly classified as smishing.

$$TP = n_{smish \rightarrow smish} \qquad (2)$$

- ***False Positive (FP)*** – FP is number of ham messages incorrectly classified as smishing.

$$FP = n_{ham \rightarrow smish} \qquad (3)$$

- ***True Negative (TN)*** – TN is number of ham messages correctly classified as ham.

$$TN = n_{ham \rightarrow ham} \qquad (4)$$

- ***False Negative (FN)*** – FN is number of smishing messages incorrectly classified as ham.

$$FN = n_{smish \rightarrow ham} \qquad (5)$$

- ***True Positive Rate (TPR)*** – It is the rate of the smishing messages that are correctly classified as smishing with respect to all existing smishing messages.

$$TPR = \frac{n_{smish \rightarrow smish}}{n_{smish \rightarrow smish} + n_{smish \rightarrow ham}} = \frac{TP}{TP + FN} \qquad (6)$$

- *False Positive Rate (FPR)* – It is the rate of the ham messages that are incorrectly classified as smishing with respect to all existing ham messages.

$$\text{FPR} = \frac{n_{ham \to smish}}{n_{ham \to ham} + n_{ham \to smish}} = \frac{FP}{TN + FP} \quad (7)$$

- *True Negative Rate (TNR)* – It is the rate of the ham messages that are correctly classified as ham with respect to all existing ham messages.

$$\text{TNR} = \frac{n_{ham \to ham}}{n_{ham \to ham} + n_{ham \to smish}} = \frac{TN}{TN + FP} \quad (8)$$

- *False Negative Rate (FNR)* – It is the rate of the smishing messages that are incorrectly classified as ham with respect to all existing smishing messages.

$$\text{FNR} = \frac{n_{smish \to ham}}{n_{smish \to smish} + n_{smish \to ham}} = \frac{FN}{FN + TP} \quad (9)$$

- *Accuracy (A)* – The effectiveness of the scheme is evaluated in term of detection accuracy which is calculated as:

$$A = \frac{TP + TN}{TP + FP + FN + T} \quad (10)$$

**Table 4.1** Classification matrix

| Classification ⇒ <br> Decision ⇓ | Smishing | Ham |
|---|---|---|
| Is Smishing | $n_{smish \to smish}$ | $n_{smish \to ham}$ |
| Is Ham | $n_{ham \to smish}$ | $n_{ham \to ham}$ |

## 4.2 Dataset Used

We have used SMS spam dataset v.1 [114] for experimental analysis. It is a publically available dataset. It consists of 5574 text messages (4,827 ham and 747 spam) in the English language. The messages are labelled as ham or spam. This dataset consists of messages from different sources. 425 spam messages are collected from the Grumbletext website [115] and 450 ham messages are extracted from Caroline Tag's Ph.D. thesis [116]. 3,375 ham messages are chosen from NUS SMS Corpus (NSC) [117]. Rest of the 1,002 ham and 322 spam messages are taken from the SMS Spam Corpus v.0.1 Big. Since smishing messages are the subset of spam messages and there is no benchmark dataset available for smishing messages, so we have manually extracted smishing messages from the spam messages. We have also gathered and included 71 smishing messages from pinterest.com [118] to our dataset. After all the processing, our final dataset consists of a total of 5169 messages, out of which 4807 are ham messages and 362 are smishing messages.

Descriptive statistics of dataset is shown in Table 4.2. After analyzing the dataset, we found that average number of characters present in ham message is 74.55, whereas 148.72 in case of smishing messages. Also, average number of words in ham message is 14.76 and 24.72 in smishing message which clearly indicates that ham messages are shorter than smishing messages. In addition, average presence of URL and symbols ($ and €) is 0.0027 and 0.0037, respectively in ham messages, whereas 0.2513 and 0.0193 in case of smishing messages which shows that probability of occurrence of URL and these symbols is more in smishing messages than in ham messages.

Table 4.2 Descriptive Statistics of the Dataset

|  | Ham Message | Smishing Message |
| --- | --- | --- |
| Total messages | 4807 | 362 |
| Average no. of character | 74.55 | 148.72 |
| Average no. of words | 14.76 | 24.72 |
| Average presence of URL | 0.0027 | 0.2513 |
| Average presence of symbols $ and € | 0.0037 | 0.0193 |

## 4.3 Results and Discussions

We implemented our scheme on a system having i5 processor, 2.4 GHz clock speed and 8 GB RAM. The backend of the entire project is developed in Python. We have designed python scripts to determine if the message is smishing message or ham message. We have used following python libraries for designing python scripts –

- Natural Language Processing Toolkit (NLTK) – It is a python library which aims at setting up Natural Language Processing (NLP) in python programming language. It provide numerous text processing libraries for classification, stemming, tokenization, parsing etc. NLTK is available for Windows, Mac OS X, and Linux.

- Comma Separated Values (CSV) – This library is used for working with data exported from spreadsheets and databases into text files formatted with fields and records, commonly referred to as comma-separated value (CSV) format because commas are often used to separate the fields in a record.
- SYS – This module provides a number of functions and variables that can be used to manipulate different parts of the python runtime environment. It contain system level information such as the version of python, and system level options such as the maximum allowed recursion depth.
- ConfigParser – This module is used for working with configuration files. It is similar to Windows INI files OS. OS modules are incorporated to provide operating system functionalities in the Python programming language.

We divided our dataset into two subsets - training dataset and testing dataset. We have used 90% of the data for training and 10% of the data for testing purpose. The training dataset consists of 4,342 ham messages and 327 smishing messages, and rest of the messages are used for testing purpose. Message and both the datasets i.e. training and testing datasets are pre-processed and normalized using algorithm 1. Thereafter, dataset is trained and message is classified as smishing or ham message using algorithm 2.

We have implemented our scheme with and without phase 1 i.e. preprocessing and normalization phase so as analyze the importance of this phase. Samples of trained ham and smishing dataset for both the cases with phase 1 and without phase 1 is shown in Table 4.3 – 4.6.

After analyzing the sample of training dataset from table 4.3 and 4.5, we have found that smishing probability of the terms that are used in smishing messages is increased after preprocessing and normalization, which in turn increases the detection accuracy of smishing messages. For example Term 'Claim' has 0.250764 smishing probability but only 0.002533 Ham probability which shows that term 'Claim' has more chances of occurrence in smishing messages. Also, earlier the smishing probability of term 'Call' was 0.443425 but after applying phase 1, this probability increased to 0.464832 as shown in table 4.3 and 4.5. Similar is the case for ham message, Ham Probability of terms used in ham messages is increased after preprocessing and normalization as shown in Table 4.4 and 4.6.

**Table 4.3** Trained Smishing Dataset without Preprocessing and Normalization

| Term | Smishing probability |
|------|----------------------|
| Call | 0.443425 |
| Bank | 0.012232 |
| Cash | 0.159021 |
| Sale | 0.006116 |
| Offer | 0.033639 |
| Prize | 0.229358 |
| Free | 0.159021 |
| Won | 0.177370 |
| Claim | 0.250764 |

**Table 4.4** Trained Ham Dataset without Preprocessing and Normalization

| Term | Ham probability |
|------|-----------------|
| Call | 0.062414 |
| Cash | 0.002303 |
| Sale | 0.001382 |
| Offer | 0.003685 |
| Free | 0.030401 |
| Won | 0.005067 |
| Claim | 0.002533 |

**Table 4.5** Trained Smishing Dataset after Preprocessing and Normalization

| Term | Smishing probability |
|------|----------------------|
| Call | 0.464832 |
| Bank | 0.015291 |
| Cash | 0.159021 |
| Sale | 0.006116 |
| Offer | 0.055046 |
| Prize | 0.232416 |
| Free | 0.159021 |
| Won | 0.17737 |
| Claim | 0.253823 |

**Table 4.6** Trained Ham Dataset after Preprocessing and Normalization

| Term | Ham probability |
|------|-----------------|
| Call | 0.071165 |
| Cash | 0.002533 |
| Sale | 0.001842 |
| Offer | 0.004376 |
| Free | 0.030631 |
| Won | 0.005067 |
| Claim | 0.002994 |

After training, we tested our scheme with the rest of the dataset. When we implemented our scheme without preprocessing and normalization, it achieves an accuracy of 88.20% with TPR of 94.28% and TNR of 87.74%. On the other hand, when we implemented our scheme with preprocessing and normalizing, it achieves an accuracy of 96.20% with TPR of 97.14% and TNR of 96.12%. The results obtained indicates that use of preprocessing and normalization phase has significantly increased the accuracy of the scheme by 8%. This shows that converting the message into its standard form can enhance the classification accuracy of the designed scheme.

The results obtained without and with Preprocessing and Normalization is shown in Table 4.7.

**Table 4.7** Experimental Results of the Scheme

|  | TPR | TNR | FPR | FNR | Accuracy |
|---|---|---|---|---|---|
| **Without Preprocessing and Normalization** | 94.28% | 87.74% | 12.25% | 5.71% | 88.20% |
| **With Preprocessing and Normalization** | 97.14% | 96.12% | 3.87% | 2.85% | 96.20% |

Comparative analysis of our approach with the other existing approaches [7, 80, 87] is shown in Table 4.8. We have also included our both considered cases in comparative analysis table and the results shows that our scheme is able to achieve highest classification accuracy and True positive rate among all the schemes discussed.

Table 4.8 Comparison of our proposed approach with the existing approaches

| Authors | Content based Analysis | Text Normalization | Algorithm Used | Advantages |
|---|---|---|---|---|
| Joo et al. [7] | ✓ | x | Naïve Bayes Algorithm | It is expected to provide security, availability, and reliability in preventing more intelligent and more malicious security threats. |
| Yadav et al. [80] | ✓ | x | Naïve Bayes Algorithm, SVM | Accuracy obtained is 84.75% with 72.5% TPR and 97% TNR. |
| Lee et al. [87] | ✓ | x | Source and content of the message is analyzed | Processing is done at Cloud environment. User involvement reduces the false detection rate. |
| Proposed model without Preprocessing and Normalization | ✓ | x | Naïve Bayes Algorithm | Accuracy obtained is 88.20% with 94.28% TPR and 87.74% TNR. |
| Proposed model with Preprocessing and Normalization | ✓ | ✓ | Naïve Bayes Algorithm | Accuracy obtained is 96.20% with 97.14% TPR and 96.12% TNR. |

## 4.4 Chapter Summary

In this chapter, we presented the implementation details of our proposed model and discussed the results. We also mentioned the dataset and platform used. We have implemented our smishing security scheme with and without preprocessing and normalization process. Results shows that pre-processing and normalization process improves the classification accuracy of the scheme from 88.20% to 96.20%. Lastly, we present a comparative analysis of our proposed approach with some of the other related solutions.

# CHAPTER 5

# CONCLUSION AND FUTURE WORK

The aim of phishing attack is to steal personal information of the user. Although phishing attack has been targeting the desktop users from a very long time, but now the attackers have shifted their focus to mobile device users. When it comes to mobile phones, the attackers have numerous ways to reach the users and some of these include SMS, mobile applications, e-mails, mobile web browsers and MMS. Social engineering is one of the most widely used methods to acquire user's information using fake websites, emails and SMSes. Fraud messages are sent to users, asking them to update their details. Malicious software is installed on device of user either by sending the malicious links or making it available on the application store. It is difficult for the users to ignore the SMS they receive on their mobile phone devices. In this dissertation, we analyze the mobile phishing attacks in a broad view, however, we restrict our work and proposed solution only to detection and prevention of smishing attacks.

## 5.1 Conclusion

With the increase in popularity of smartphone devices, threats related to these devices have also increased. Smishing attack is one such threat and it has become one of the major issue faced by the mobile phone users. Identification of smishing attack with higher accuracy is an important research issue as not much amount of work has been done in this field. Various solutions have been proposed for detecting smishing attacks but still there is lack of complete solution. One of the major challenge faced by the researchers is the presence of abbreviations, short forms and slang words in the text message due to which a term can be written and interpreted in various different ways. In addition, due to the short length of the text message, very less number of features can be extracted from the message. Moreover, it is important to maintain the privacy of the user as SMS may contain personal information that should not be revealed to the third party.

To address these problems, we have proposed a smishing security model. In the first phase, the message is pre-processed and then normalized with the help of NoSlang dictionary. It is done to replace the message by its standard form. In the second phase, Bayesian learning is used to train the dataset and after training, a ham and smishing training dataset is created. Naïve Bayes classifier is used to determine if the message is smishing message or ham message. We evaluated our approach with the publically available dataset. After experimental analysis, we found that preprocessing and normalization process significantly increases the classification accuracy of the scheme from 88.20% to 96.20%. The proposed scheme achieves 96.20% classification accuracy with 97.14% TPR and 96.12% TNR.

## 5.2 Future Work

Several approaches have been proposed in order to protect users from SMS Phishing attacks, but still the threat is not elevated and demands more attention towards the improvement of defense solutions. Even though the results obtained from the experiments are satisfactory, but attackers always find new ways to trick users. Future work that can be done in the proposed mechanism includes –

- URL present in the message can be analysed so as to determine if this URL redirects user to a malicious login page or leads to download of some malicious application.
- Normalization process can be improved so that words of the message are expanded as per the context of the message from among all the available concepts related to that word.
- Size of the dataset can be increased in order to increase the richness of the scheme.

# Publications out of research work

1. Diksha Goel and Ankit Kumar Jain, (2018). Mobile phishing attacks and defence mechanisms: state of art and open research challenges. *Computers & Security,* Elsevier, vol 73, pp. 519-544, DOI: 10.1016/j.cose.2017.12.006. **(SCI- Indexed) (Impact Factor: 2.849)**

2. Diksha Goel and Ankit Kumar Jain, (2017). Smishing-Classifier: A Novel Framework for detection of Smishing Attack in Mobile Environment. *In proceedings of International Conference on Next Generation Computing Technologies,* Dehradun, India, pp. 502-512. DOI: 10.1007/978-981-10-8660-1_38. **(Scopus- Indexed)**

3. Diksha Goel and Ankit Kumar Jain, (2017). Overview of Smartphone Security- Attack and Defence Techniques. *Computer and Cyber Security: Principles, Algorithm, Applications and Perspectives*, CRC Press, Taylor & Francis. (Accepted, In press) **(Book Chapter)**

4. Diksha Goel and Ankit Kumar Jain, (2018). A Content based Approach for Detecting Smishing Attack in Mobile Environment. *Journal of Ambient Intelligence & Humanized Computing (AIHC)*, Springer. (Under Review) **(SCI- Indexed) (Impact Factor: 1.588)**